\documentclass[prd,nopacs,preprintnumbers,twocolumn,amsmath,nofootinbib,amssymb]{revtex4}
\usepackage{graphicx,color,dcolumn,booktabs,bm}
\usepackage{longtable,lscape}
\usepackage{txfonts}
\usepackage{overpic}
\usepackage{epstopdf}
\usepackage{indentfirst}
\usepackage{feynmf}   
\usepackage{slashed}  
\usepackage{cases}
\usepackage{color}
\usepackage{float}
\usepackage{makecell}
\usepackage{multirow}
\usepackage{graphicx,color,dcolumn,booktabs,bm}
\usepackage{epsfig,dsfont,amssymb,amsmath,amsfonts,amsbsy,mathrsfs}

\graphicspath{{Figures/}} %

\usepackage{hyperref}
\hypersetup{colorlinks,citecolor=blue,anchorcolor=red,menucolor=red, linkcolor=red,filecolor=red,runcolor=red,urlcolor=blue,frenchlinks=red}

\makeatletter
\@addtoreset{equation}{section}
\makeatother

\allowdisplaybreaks

\newcolumntype{I}{!{\vrule width 0.9pt}}

\begin{document}

\title{Investigating the transition form factors of $\Lambda_b\to\Lambda_c(2625)$ and $\Xi_b\to\Xi_c(2815)$ and the corresponding weak decays with support from baryon spectroscopy}
\author{Yu-Shuai Li$^{1,2}$}\email{liysh20@lzu.edu.cn}
\author{Xiang Liu$^{1,2,3}$}\email{xiangliu@lzu.edu.cn}
\affiliation{$^1$School of Physical Science and Technology, Lanzhou University, Lanzhou 730000, China\\
$^2$Research Center for Hadron and CSR Physics, Lanzhou University and Institute of Modern Physics of CAS, Lanzhou 730000, China\\
$^3$Lanzhou Center for Theoretical Physics, Key Laboratory of Theoretical Physics of Gansu Province, and Frontiers Science Center for Rare Isotopes, Lanzhou University, Lanzhou 730000, China}

\begin{abstract}
We calculate the form factors of the $\Lambda_b\to\Lambda_c(2625)$ and $\Xi_b\to\Xi_c(2815)$ transitions, and additionally evaluate the corresponding semileptonic decays and the color-allowed two-body nonleptonic decays. In order to obtain the concerned form factors, we use the three-body light-front quark model with the support from baryon spectroscopy. In this work, as important physical inputs, the spatial wave functions of concerned baryons are obtained by the Gaussian expansion method with a semirelativistic potential model. For the semileptonic processes, the branching ratios of the electron and muon channels can reach up to the order of magnitude of $1\%$, where our result $\mathcal{B}(\Lambda_b^0\to\Lambda_c^+(2625)\mu^-\nu_{\mu})=(1.641\pm0.113)\%$ is consistent with the current experimental data. As for the nonleptonic processes, the decays to $\pi^-$, $\rho^-$, and $D_s^{(*)-}$ final states have considerable widths. These discussed decay modes could be accessible at the LHCb experiment.
\end{abstract}

\maketitle

\section{introduction}
\label{sec1}

The investigation of bottom baryon weak decay is a fiery topic in heavy flavor physics, which has drawn attentions at both theoretical and experimental arenas.
{There exist abundant decay modes involved in bottom baryons due to their higher mass.}
So the bottom baryon decay provides a superb platform to test the Quantum Chromodynamics (QCD), and to search for new physics beyond the Standard Model (SM) via detecting whether the lepton flavor universality (LFU) is violated or not~\cite{BaBar:2012obs,BaBar:2013mob,Belle:2015qfa,LHCb:2015gmp,Belle:2016dyj,Belle:2019rba,FermilabLattice:2021cdg}. Besides, it is also helpful to discover the new exotic states including the hidden-charm pentaquark states $P_{c}(4312)$, $P_{c}(4380)$, $P_{c}(4440)$, and $P_{c}(4457)$ in $\Lambda_b\to J/\psi pK$~\cite{LHCb:2015yax,LHCb:2019kea} mode, the $P_c(4337)$ in the $B_s^0\to J/\psi p\bar{p}$~\cite{LHCb:2021chn} mode,  and the $P_{cs}(4459)$ in the $\Xi_b\to J/\psi\Lambda K$~\cite{LHCb:2020jpq} mode. 

In theoretical aspect, the bottom baryons decaying into the $J^{P}=1/2^{+}$ ground charmed baryons by both semileptonic and nonleptonic processes have been widely studied via various theoretical approaches, including lattice QCD (LQCD)~\cite{Gottlieb:2003yb,Detmold:2015aaa}, QCD sum rules~\cite{Huang:2005mea,Azizi:2018axf,Zhao:2020mod}, light-cone sum rules~\cite{Wang:2009yma,Duan:2022uzm,Miao:2022bga}, and various phenomenological quark models~\cite{Pervin:2005ve,Ebert:2006rp,Ke:2007tg,Gutsche:2015mxa,Faustov:2016pal,Gutsche:2018nks,Chua:2018lfa,Ke:2019smy,Chua:2019yqh,Rahmani:2020kjd,Geng:2020ofy,Li:2021qod,Li:2021kfb}. However, compared with these studies mentioned above, the investigation of the decays of bottom baryon into the $P$-wave baryon state should still be paid more attention. In the past years, some theoretical groups were dedicated to this issue. For example, Pervin {\it et al.} studied the semileptonic decays of $\Lambda_{b}$ into $\Lambda_c$ baryon with $J^{P}=(1/2^{\pm},3/2^{-})$ by a constituent quark model with both nonrelativistic and semirelativistic Hamiltonians~\cite{Pervin:2005ve}. Gutsche {\it et al.} also studied the same channels by a covariant confined quark model (CCQM)~\cite{Gutsche:2018nks}. The heavy quark spin symmetry (HQSS) was also applied to estimate the semileptonic decays $\Lambda_b\to\Lambda_c(2595,2625)\ell^{-}\nu_{\ell}$~\cite{Nieves:2019kdh}. Besides, Meinel and Rendon performed the first LQCD calculation of the $\Lambda_b\to\Lambda_c(2595,2625)\ell^{-}\nu_{\ell}$ decays~\cite{Meinel:2021rbm,Meinel:2021mdj}. For the nonleptonic process, Chua calculated a series of color-allowed decay of bottom baryon into $P$-wave charmed baryon by the light-front quark model (LFQM)~\cite{Chua:2019yqh}. And Liang {\it et al.} evaluated the nonleptonic $\Lambda_b\to\Lambda_c(2595,2625)\pi^{-}$ \cite{Liang:2016ydj}, $\Lambda_b\to\Lambda_c(2595,2625)D_{s}^{-}$ \cite{Liang:2016ydj}, and semileptonic $\Lambda_b\to\Lambda_c(2625)\ell\nu_{\ell}$~\cite{Liang:2016exm} decays by assuming the $\Lambda_c(2595)$ and $\Lambda_c(2625)$ as a dynamically generated resonance from the $DN$, $D^{*}N$ interaction and coupled channels~\cite{Liang:2016ydj,Liang:2016exm}. Pavao {\it et al.} carried out the investigation of the $\Xi_b^-\to\Xi_c^0(2815)\pi^{-}(D_{s}^{-})$ and $\Xi_b^-\to\Xi_c^0(2815)\ell^{-}\nu_{\ell}$ processes when treating the $\Xi_c(2815)$ as dynamically generated resonance  from the vector meson-baryon interaction~\cite{Pavao:2017cpt}.

In our former work, we once studied the form factors and the semileptonic decays into charmed baryons with $J^{P}=1/2^{-}$ by LFQM~\cite{Li:2021qod}, which is supported by the baryon spectroscopy. This treatment is
 different from that given in Ref.~\cite{Chua:2019yqh}. We should indicate that there exists difference of the results of these discussed transition when adopting different frameworks~\cite{Pervin:2005ve,Liang:2016ydj,Liang:2016exm,Pavao:2017cpt,Gutsche:2018nks,Nieves:2019kdh,Chua:2019yqh}, which should be clarified by further experimental measurement. In general, the issue around these weak transitions of bottom baryons is still open.

{As a continuation of the decays into charmed baryons with
$J^P = 1/2^-$~\cite{Li:2021qod}}, in this work, we investigate the weak transitions relevant to the $J^{P}=3/2^{-}$ charmed baryons, which include the $\Lambda_b\to\Lambda_c(2625)$ and $\Xi_b\to\Xi_c(2815)$ processes, where the $\Lambda_c(2625)$ and $\Xi_c(2815)$ are treated as the conventional $\rho$-mode excited $P$-wave charmed baryons\footnote{{In this work, we do not consider several other possible spin 3/2
$\Lambda_c$ and $\Xi_c$ resonances such as the $\Lambda_c(2860)$ \cite{LHCb:2017jym} and $\Xi_c(2645)$ \cite{Belle:2016lhy}, which have positive parity different from that of the $\Lambda_c(2625)$ and $\Xi_c(2815)$.
Here, the $\Lambda_c(2860)$ and $\Xi_c(2645)$ are good candidates of $D$-wave \cite{Chen:2017aqm} and $S$-wave charmed baryons \cite{Chen:2016iyi}, while the discussed $\Lambda_c(2625)$ and $\Xi_c(2815)$ are assigned as $P$-wave charmed baryons.}}. Note that this assignment is suitable since the experimental mass value of the $\Lambda_c(2625)$ and $\Xi_c(2815)$ can be reproduced by the potential models~\cite{Chen:2015kpa,Guo:2019ytq,Yu:2022ymb,Li:2022xtj,Capstick:1985xss}. {Although we
adopt the similar way given by Ref. \cite{Li:2021qod}, we still want to emphasize the improvement made in this work. First of all, the involved charmed baryons in the final state have $3/2^-$ quantum number, which makes whole deduction framework become more complicated. Especially, at present the study around the production of charmed baryons with $3/2^-$ is not enough compared with that relevant to these low-lying charmed baryons. Our work is a timely investigation of this issue.
Secondly, in Ref. \cite{Li:2021qod}, we focus on the weak decays of the $\Lambda_b$ baryon into $1/2^\pm$ charmed baryons. However, in the present work we study the $\Xi_b$ decays into $3/2^-$ charmed baryons, which is motivated by the experiment fact that the data of the $\Xi_b$ bottom baryon can be also largely produced in the $pp$ collisions at Large Hadron Collider (LHC) \cite{LHCb:2019sxa}. Obviously, the present work is just at the right time, which may provide valuable hint to future experimental search for these discussed decays. Especially,
with the high-luminosity upgrade to LHC, the LHCb experiment will have great potential to explore these discussed transitions.}

As indicated in Ref. \cite{Li:2021qod}, the baryon spectroscopy can provide important input to the
spatial wave functions of these involved baryons when estimating the weak transition matrix element or the corresponding form factors. In the realistic calculation of baryon spectroscopy, we adopt the three quarks treatment, which is different from the quark-diquark approximation used by former theoretical works of weak decays~\cite{Ke:2007tg,Wang:2017mqp,Zhao:2018zcb,Zhao:2018mrg,Chua:2018lfa,Ke:2017eqo,Zhu:2018jet,Chua:2019yqh,Wang:2022ias,Zhao:2022vfr}. With the support from baryon spectroscopy, the dependence of the result on the $\beta$ value, which is a parameter of the simple hadronic oscillator, can be avoided as indicated in Refs.~\cite{Li:2021qod,Li:2021kfb}. In the next section, we will introduce more details of the deduction.

This paper is organized as follows. After the Introduction, the deduction of the formulas of eight transition form factors of the $\mathcal{B}_b(1/2^+)\to\mathcal{B}_c(3/2^-)$ process is given in Sec.~\ref{sec2}. For obtaining the spatial wave functions of the involved baryons, we introduce the semirelativistic potential model and adopt the Gaussian expansion method (GEM) in Sec.~\ref{sec3}. And then, in Sec.~\ref{sec4} we present the results of the form factors of the $\Lambda_b\to\Lambda_c(2625)$ and $\Xi_b\to\Xi_c(2815)$ transitions, and further evaluate the corresponding semileptonic decays and color-allowed two-body nonleptonic decays. Finally, this paper ends with the discussion and conclusion.

\section{The transition form factors of the bottom baryon to the charmed baryon}
\label{sec2}

The $b\to c$ weak decay is usually dependent on the hadronic structure reflected by the baryon to baryon weak transition matrix element $\langle\mathcal{B}_{c}\vert\bar{c}\gamma^{\mu}(1-\gamma^{5})b\vert\mathcal{B}_{b}\rangle$. In this section, we briefly introduce how to calculate the matrix element. Since the constituent quarks are confined in hadron, the matrix element cannot be calculated by the perturbative QCD. Usually, the matrix element can be parameterized in terms of a series of dimensionless form factors~\cite{Chua:2019yqh}, i.e.,
\begin{equation}
\begin{split}
\langle\mathcal{B}_{c}(&3/2^-)\vert\bar{c}\gamma^{\mu}b\vert\mathcal{B}_{b}(1/2^+)\rangle\\
=&\bar{u}_{\alpha}(P^{\prime},J_z^{\prime})\bigg{[}
g_1^V(q^2)g^{\alpha\mu}
+g_2^V(q^2)\frac{P^{\alpha}}{M}\gamma^{\mu}+g_3^V(q^2)\frac{P^{\alpha}P^{\prime\mu}}{MM^{\prime}}\\
&+g_4^V(q^2)\frac{P^{\alpha}P^{\mu}}{M^2}
\bigg{]}u(P,J_z),
\label{eq:ffsV}
\end{split}
\end{equation}
\begin{equation}
\begin{split}
\langle\mathcal{B}_{c}(&3/2^-)\vert\bar{c}\gamma^{\mu}\gamma^{5}b\vert\mathcal{B}_{b}(1/2^+)\rangle\\
=&\bar{u}_{\alpha}(P^{\prime},J_z^{\prime})\bigg{[}
f_1^A(q^2)g^{\alpha\mu}
+f_2^A(q^2)\frac{P^{\alpha}}{M}\gamma^{\mu}
+f_3^A(q^2)\frac{P^{\alpha}P^{\prime\mu}}{MM^{\prime}}\\
&+f_4^A(q^2)\frac{P^{\alpha}P^{\mu}}{M^2}
\bigg{]}\gamma^{5}u(P,J_z),
\label{eq:ffsA}
\end{split}
\end{equation}
where $M$ and $M^{\prime}$ are the masses of the initial bottom baryon and the final charmed baryon, respectively. $P$ and $P^{\prime}$ are the corresponding three-momentum, and $J_z$ and $J_z^{\prime}$ are the third components of the spins. Here, we ignore the spin quantum number since they are definite.

{ In this work, we use the standard light-front quark model to calculate the relevant form factors. The light-front quark model, which was proposed by Terentev and Berestetsky in a relativistic quark model \cite{Terentev:1976jk,Berestetsky:1977zk} based on the light front formalism and light front quantization of QCD, have been widely and successfully used in studying the weak decay form factors (see Ref. \cite{Chang:2019obq} and its references). In this work, we take the same framework \cite{Ke:2019smy,Ke:2019lcf,Ke:2021pxk} to calculate the relevant form factors.
In the concrete calculation, there exists input of the spatial wave function of these discussed baryon states. Usually, one takes a simple Harmonic Oscillator (SHO) wave function, which must result in the calculated physical quantities dependent on the $\beta$ value, which is the parameter in the SHO wave function. For avoiding such problem, we proposed to directly adopt the numerical spatial wave function by solving the potential model with the help of the Gaussian expansion method \cite{Li:2021qod,Li:2021kfb}.}
In analog to Refs. \cite{Cheung:1995ub,Cheng:1996if,Geng:1997ws,Cheng:2004cc,Wang:2017mqp,Ke:2019smy,Ke:2019lcf,Geng:2022xpn}, the vertex function of a single heavy flavor baryon $\mathcal{B}_{Q}$ with spin $J$ and momentum $P$ can be written as
\begin{equation}
\begin{split}
\vert\mathcal{B}_{Q}(P,J&,J_z)\rangle=\int\frac{d^3\tilde{p}_{1}}{2(2\pi)^3}\frac{d^3\tilde{p}_{2}}{2(2\pi)^3}\frac{d^3\tilde{p}_{3}}{2(2\pi)^3}2(2\pi)^3\\
&\times\sum_{\lambda_1,\lambda_2,\lambda_3}\Psi^{J,J_z}(\tilde{p}_i,\lambda_i)C^{\alpha\beta\gamma}
\delta^{3}(\tilde{P}-\tilde{p}_1-\tilde{p}_2-\tilde{p}_3)\\
&\times~F_{q_1q_2Q}~\vert q_{1\alpha}(\tilde{p}_1,\lambda_1)\rangle~\vert q_{2\beta}(\tilde{p}_2,\lambda_2)\rangle~\vert Q_{\gamma}(\tilde{p}_3,\lambda_3)\rangle,
\end{split}
\end{equation}
where $C^{\alpha\beta\gamma}$ and $F_{q_1q_2q_3}$ represent the color and flavor factors, respectively, and $\lambda_i$ and $p_i$ ($i$=1,2,3) are the helicities and light-front momenta of the on-mass-shell quarks, respectively, defined as
\begin{equation}
\tilde{p}_i=(p_i^+, \vec{p}_{i\bot}), \quad p_i^+=p_i^0+p_i^3, \quad \vec{p}_{i\bot}=(p_i^1, p_i^2).
\end{equation}
For describing the motions of the constituents, we should introduce the intrinsic variables $(x_{i},~\vec{k}_{i})$ ($i=1,2,3$)
\begin{equation}
p_{i}^{+}=x_{i}^{}P^{+},~~
\vec{p}_{i\bot}=x_{i}\vec{P}_{i\bot}+\vec{k}_{i\bot},~~
\sum_{i=1}^{3}\vec{k}_{i\bot}=0,~~
\sum_{i=1}^{3}x_{i}=1,
\end{equation}
where $x_{i}$ are the light-front momentum fractions constrained by $0<x_{i}<1$.

In this work, the spin-spatial wave functions for anti-triplet single heavy baryon $\mathcal{B}_{Q}(\bar{3}_f,J^P=1/2^+)$ and $\mathcal{B}_{Q}(\bar{3}_f,J^P=3/2^-)$ are written as~\cite{Korner:1994nh,Hussain:1995xs,Tawfiq:1998nk}
\begin{equation}
\begin{split}
\Psi^{1/2,J_z}(\tilde{p}_i,\lambda_i)=&A_0\bar{u}(p_1,\lambda_1)[(\slashed{P}+M_0)\gamma^5]v(p_2,\lambda_2)\\
&\times\bar{u}_{Q}(p_3,\lambda_3)u(P,J_z)\psi(x_i,\vec{k}_{i}),\\
\Psi^{3/2,J_z}(\tilde{p}_i,\lambda_i)=&B_0\bar{u}(p_1,\lambda_1)[(\slashed{P}+M_0)\gamma^5]v(p_2,\lambda_2)\\
&\times\bar{u}_{Q}(p_3,\lambda_3)K^{\alpha}u_{\alpha}(P,J_z)\psi(x_i,\vec{k}_{i}),
\label{eq:SpinSpatialWaveFunction}
\end{split}
\end{equation}
respectively.

As the fundamental inputs, the spatial wave functions $\psi$ should be discussed here. Usually, the single heavy flavor baryon is regarded as a quasi two-body bound state of the light quark cluster with heavy quark ($b\ (\text{or}\ c)$) to form the $\rho$-mode excitation. {The spatial wave function of a single heavy baryon can be written as~\cite{Ke:2019smy,Ke:2019lcf,Ke:2021pxk}
\begin{equation}
\begin{split}
\psi(x_i,\vec{k}_{i})=&N_{\psi}\sqrt{\frac{e_1e_2e_3}{x_1x_2x_3M_{0}}}\phi_{\rho}\Big{(}\frac{m_1\vec{k}_{2}-m_2\vec{k}_{1}}{m_1+m_2}\Big{)}\\
&\times\phi_{\lambda}\Big{(}\frac{(m_1+m_2)\vec{k}_{3}-m_3(\vec{k}_{1}+\vec{k}_{2})}{m_1+m_2+m_3}\Big{)},
\label{eq:SpatialWaveFunction}
\end{split}
\end{equation}
where $\vec{k}_{i}=(\vec{k}_{i\bot},k_{iz})$ with \cite{Ke:2019smy}
\begin{equation}
k_{iz}=\frac{x_{i}M_{0}}{2}-\frac{m_{i}^{2}+\vec{k}_{i\bot}^{2}}{2x_{i}M_{0}}.
\end{equation}
The $\phi_{\rho(\lambda)}$ is the spatial wave function of $\rho(\lambda)$-mode excitation.}

The normalized factors in Eq.~\eqref{eq:SpinSpatialWaveFunction} are expressed as
\begin{equation*}
\begin{split}
A_0&=\frac{1}{\sqrt{16P^+M_0^3(e_1+m_1)(e_2+m_2)(e_3+m_3)}},\\
B_0&=\frac{\sqrt{3}}{\sqrt{16P^+M_0^3(e_1+m_1)(e_2+m_2)(e_3-m_3)(e_3+m_3)^2}},
\end{split}
\end{equation*}
where the factor in Eq.~\eqref{eq:SpatialWaveFunction} is $N_{\psi}=(4\pi^{3/2})^2$ for the ground state and $N_{\psi}=(4\pi^{3/2})^2/\sqrt{3}$ for the $P$-wave state. These factors are determined by the following normalizations:
\begin{equation}
\begin{split}
\sum_{J_z,J_z^{\prime}}&\langle\mathcal{B}_{Q}(P^{\prime},J,J_z^{\prime})\vert\mathcal{B}_{Q}(P,J,J_z)\rangle
=\sum_{J_z,J_z^{\prime}}2(2\pi)^3P^{+}\delta^3(\tilde{P}-\tilde{P}^{\prime})\delta_{J_z,J_z^{\prime}},
\label{eq:normalization1}
\end{split}
\end{equation}
and
\begin{equation}
\begin{split}
\int&\Bigg{(}\prod_{i=1}^{3}\frac{dx_id^2\vec{k}_{i\bot}}{2(2\pi)^3}\Bigg{)}2(2\pi)^3\delta\Big{(}1-\sum_{i}x_i\Big{)}\\
&\times\delta^2\Big{(}\sum_{i}\vec{k}_{i\bot}\Big{)}\psi^{*}(x_i,\vec{k}_{i})\psi(x_i,\vec{k}_{i})=1.
\label{eq:normalization2}
\end{split}
\end{equation}

With the above vertex wave functions in the framework of LFQM, the general expression of the weak transition matrix element can be expressed as
\begin{widetext}
\begin{equation}
\begin{split}
\langle\mathcal{B}_{c}(P^{\prime},J_z^{\prime})\vert\bar{c}\Gamma^{\mu}_{i}b\vert\mathcal{B}_{b}(P,J_z)\rangle
=&\int\Big{(}\frac{dx_1d^2\vec{k}_{1\bot}}{2(2\pi)^3}\Big{)}
\Big{(}\frac{dx_2d^2\vec{k}_{2\bot}}{2(2\pi)^3}\Big{)}
\frac{\psi_c^{\ast}(x_i^{\prime},\vec{k}_{i\bot}^{\prime})\psi_b(x_i,\vec{k}_{i\bot})}{(16/\sqrt{3})\sqrt{x_3x_3^{\prime}M_0^3M_0^{\prime3}}}\\
&\times\frac{\text{Tr}[(\slashed{P}^{\prime}-M_0^{\prime})\gamma^5(\slashed{p}_1+m_1)(\slashed{P}+M_0)\gamma^5(\slashed{p}_2-m_2)]}
{\sqrt{(e_1+m_1)(e_2+m_2)(e_3+m_3)
(e_1^\prime+m_1^\prime)(e_2^\prime+m_2^\prime)(e_3^\prime-m_3^\prime)(e_3^{\prime}+m_3^{\prime})^2}}\\
&\times\bar{u}_{\alpha}(P^{\prime},J_z^{\prime})K^{\prime\alpha}(\slashed{p}_3^{\prime}+m_3^{\prime})\Gamma^{\mu}_{i}(\slashed{p}_3+m_3)u(P,J_z).
\label{eq:ffs}
\end{split}
\end{equation}
Here, the Lorentz structures is defined as $\Gamma^{\mu}_i=\big{\{}\gamma^{\mu},\gamma^{\mu}\gamma^{5}\big{\}}$, $K^{\prime}=\big{[}(m_1^{\prime}+m_2^{\prime})p_3^{\prime}-m_3^{\prime}(p_1^{\prime}+p_2^{\prime})\big{]}/\big{(}m_1^{\prime}+m_2^{\prime}+m_3^{\prime}\big{)}$ is the $\lambda$-mode momentum of the $P$-wave charmed baryon, and the $\psi_{b}$ and $\psi_{c}$ are the spatial wave functions of the bottom baryon and the charmed baryon, respectively.

Next, we should introduce how to extract the form factors by setting the $q^+=0$ and $\vec{q}_{\bot}\neq0$ condition. To extract the four form factors of the vector current, we multiply $\bar{u}(P,J_z)\Gamma_{i}^{V,\mu\beta}u_{\beta}(P^{\prime},J_z^{\prime})$ on both sides of Eq.~\eqref{eq:ffs} with specifically setting $\Gamma_i^{\mu}=\gamma^{\mu}$, and then sum over the polarizations of the initial and the final baryons. The left side can be replaced by Eq.~\eqref{eq:ffsV}, and the right side can be calculated by performing the traces and then the integrations. The Lorentz structures are chosen as $\Gamma_{i}^{V,\mu\beta}=\big{\{}g^{\beta\mu},P^{\beta}\gamma^{\mu},P^{\beta}P^{\prime\mu},P^{\beta}P^{\mu}\big{\}}$~\cite{Wang:2022ias,Zhao:2022vfr}. The complete expressions of the form factors of the vector current are
\begin{equation}
\begin{split}
g_1^{V}(q^2)=&-\frac{1}{2\tilde{Q}_+}G_1^{V}(q^2)
-\frac{M_0^\prime}{2\tilde{Q}_{-}\tilde{Q}_{+}}G_2^{V}(q^2)
+\frac{M_0^2+M_0M_0^\prime+M_0^{\prime2}-q^2}{\tilde{Q}_{-}\tilde{Q}_{+}^2}G_3^{V}(q^2)
-\frac{M_0^{\prime2}}{\tilde{Q}_{-}\tilde{Q}_{+}^2}G_4^{V}(q^2),\\
g_2^{V}(q^2)=&-\frac{MM_0^\prime}{2\tilde{Q}_{-}\tilde{Q}_{+}}G_1^{V}(q^2)
-\frac{2MM_0^{\prime2}}{\tilde{Q}_{-}^2\tilde{Q}_{+}}G_2^{V}(q^2)
+\frac{MM_0^\prime(M_0^2+4M_0M_0^\prime+M_0^{\prime2}-q^2)}{\tilde{Q}_{-}^2\tilde{Q}_{+}^2}G_3^{V}(q^2)
+\frac{2MM_0^{\prime3}}{\tilde{Q}_{-}^2\tilde{Q}_{+}^2}G_4^{V}(q^2),\\
g_3^{V}(q^2)=&\frac{MM^\prime(M_0^2+M_0M_0^\prime+M_0^{\prime2}-q^2)}{\tilde{Q}_{-}\tilde{Q}_{+}^2}G_1^{V}(q^2)
+\frac{MM^{\prime}M_{0}^{\prime}(M_0^2+4M_0M_0^\prime+M_0^{\prime2}-q^2)}{\tilde{Q}_{-}^2\tilde{Q}_{+}^2}G_2^{V}(q^2)\\
&-\frac{2MM^{\prime}(M_0^4+2M_0^3M_0^{\prime}+2M_0M_0^{\prime}(M_0^{\prime2}-q^2)+(M_0^{\prime2}-q^2)^2+2M_0^2(6M_0^{\prime2}-q^2))}{\tilde{Q}_{-}^2\tilde{Q}_{+}^3}G_3^{V}(q^2)\\
&+\frac{4MM^{\prime}M_0^{\prime2}(2M_0^2-M_0M_0^{\prime}+2(M_0^{\prime2}-q^2))}{\tilde{Q}_{-}^2\tilde{Q}_{+}^3}G_4^{V}(q^2),\\
g_4^{V}(q^2)=&-\frac{M^2M_0^{\prime2}}{\tilde{Q}_{-}\tilde{Q}_{+}^2}G_1^{V}(q^2)
+\frac{2M^2M_0^{\prime3}}{\tilde{Q}_{-}^2\tilde{Q}_{+}^2}G_2^{V}(q^2)
+\frac{4M^2M_0^{\prime2}(2M_0^2-M_0M_0^{\prime}+2(M_0^{\prime2}-q^2))}{\tilde{Q}_{-}^2\tilde{Q}_{+}^3}G_3^{V}(q^2)
-\frac{20M^2M_0^{\prime4}}{\tilde{Q}_{-}^2\tilde{Q}_{+}^3}G_4^{V}(q^2),
\end{split}
\end{equation}
where $M$ and $M^{\prime}$ are the physical masses of the bottom and charmed baryons, respectively, and $\tilde{Q}_{\pm}=(M_0\pm M_0^\prime)^2-q^2$ with
\begin{equation}
M_{0}^{(\prime)2}=\frac{\vec{k}_{1\bot}^{(\prime)2}+m_{1}^{(\prime)2}}{x_1}
+\frac{\vec{k}_{2\bot}^{(\prime)2}+m_{2}^{(\prime)2}}{x_2}
+\frac{\vec{k}_{3\bot}^{(\prime)2}+m_{3}^{(\prime)2}}{x_3}
\end{equation}
being the invariant mass square~\cite{Ke:2019smy}. Besides,
\begin{equation}
\begin{split}
G_{(1,2,3,4)}^V(q^2)=&\int\bigg{(}\frac{dx_1d^2\vec{k}_{1\bot}}{2(2\pi)^3}\bigg{)}\bigg{(}\frac{dx_2d^2\vec{k}_{2\bot}}{2(2\pi)^3}\bigg{)}
\frac{\psi_b(x_i,\vec{k}_{i\bot})\psi_c^{\ast}(x_i^{\prime},\vec{k}^{\prime}_{i\bot})}{\sqrt{x_3x_3^{\prime}}}A_0B_0^{\prime}\text{Tr}[\cdots]\\
&\times\text{Tr}\big{[}(G_{\mathcal{B}_c})_{\beta\alpha}K^{\prime\alpha}(\slashed{p}_{3}^{\prime}+m_{3}^{\prime})\gamma^{\mu}(\slashed{p}_{3}+m_{3})(\slashed{P}+M_0)\Gamma_{(1,2,3,4),\mu}^{V,\beta}\big{]}
\label{eq:formfactorsV}
\end{split}
\end{equation}
with
\begin{eqnarray}
A_{0}&=&1\big{/}{\sqrt{16M_0^3(e_1+m_1)(e_2+m_2)(e_3+m_3)}},\\
B_{0}^{\prime}&=&\sqrt{3}\big{/}{\sqrt{16M_{0}^{\prime3}(e_1^\prime+m_1^\prime)(e_2^\prime+m_2^\prime)(e_3^\prime-m_3^\prime)(e_3^{\prime}+m_3^{\prime})^2}},\\
\label{eq:trace}
\text{Tr}[\cdots]&=&\text{Tr}[(\slashed{P}^{\prime}-M_0^{\prime})\gamma^5(\slashed{p}_1+m_1)(\slashed{P}-M_0)\gamma^5(\slashed{p}_2-m_2)],\\
(G_{\mathcal{B}_c})^{\mu\nu}&=&-(\slashed{P}^{\prime}+M_0^{\prime})
\Big{[}g^{\mu\nu}-\frac{1}{3}\gamma^{\mu}\gamma^{\nu}-\frac{2}{3M_0^{\prime2}}P^{\prime\mu}P^{\prime\nu}-\frac{1}{3M_0^{\prime}}\big{(}\gamma^{\mu}P^{\prime\nu}-\gamma^{\nu}P^{\prime\mu}\big{)}\Big{]}.
\end{eqnarray}
Analogously, the form factors of the axial-vector current can be extracted with the structures $\bar{u}(P,J_z)\Gamma^{A,\mu\beta}_{i}u_{\beta}(P^{\prime},J_z^{\prime})$, where $\Gamma_{i}^{A,\mu\beta}=\big{\{}g^{\beta\mu}\gamma^{5},P^{\beta}\gamma^{\mu}\gamma^{5},P^{\beta}P^{\prime\mu}\gamma^{5},P^{\beta}P^{\mu}\gamma^{5}\big{\}}$ is defined. The complete expressions of the form factors of the axial-vector current are expressed as
\begin{equation}
\begin{split}
f_1^{A}(q^2)=&\frac{1}{2\tilde{Q}_{-}}F_1^{A}(q^2)
-\frac{M_0^{\prime}}{2\tilde{Q}_{-}\tilde{Q}_{+}}F_2^{A}(q^2)
-\frac{M_0^2-M_0M_0^{\prime}+M_0^{\prime2}-q^2}{\tilde{Q}_{-}^2\tilde{Q}_{+}}F_3^{A}(q^2)
+\frac{M_0^{\prime2}}{\tilde{Q}_{-}^2\tilde{Q}_{+}}F_4^{A}(q^2),\\
f_2^{A}(q^2)=&\frac{MM_0^{\prime}}{2\tilde{Q}_{-}\tilde{Q}_{+}}F_1^{A}(q^2)
-\frac{2MM_0^{\prime2}}{\tilde{Q}_{-}\tilde{Q}_{+}^2}F_2^{A}(q^2)
-\frac{MM_{0}^{\prime}(M_0^2-4M_0M_0^{\prime}+M_0^{\prime2}-q^2)}{\tilde{Q}_{-}^2\tilde{Q}_{+}^2}F_3^{A}(q^2)
-\frac{2MM_0^{\prime3}}{\tilde{Q}_{-}^2\tilde{Q}_{+}^2}F_4^{A}(q^2),\\
f_3^{A}(q^2)=&-\frac{MM^{\prime}(M_0^2-M_0M_0^{\prime}+M_0^{\prime2}-q^2)}{\tilde{Q}_{-}^2\tilde{Q}_{+}}F_1^{A}(q^2)
+\frac{MM^{\prime}M_0^{\prime}(M_0^2-4M_0M_0^{\prime}+M_0^{\prime2}-q^2)}{\tilde{Q}_{-}^2\tilde{Q}_{+}^2}F_2^{A}(q^2)\\
&+\frac{2MM^{\prime}(M_0^4-2M_0^3M_0^{\prime}-2M_0M_0^{\prime}(M_0^{\prime2}-q^2)+(M_0^{\prime2}-q^2)^2+2M_0^2(6M_0^{\prime2}-q^2))}{\tilde{Q}_{-}^3\tilde{Q}_{+}^2}F_3^{A}(q^2)\\
&-\frac{4MM^{\prime}M_0^{\prime2}(2M_0^{2}+M_0M_0^{\prime}+2(M_0^{\prime2}-q^2))}{\tilde{Q}_{-}^3\tilde{Q}_{+}^2}F_4^{A}(q^2),\\
f_4^{A}(q^2)=&\frac{M^2M_0^{\prime2}}{\tilde{Q}_{-}^2\tilde{Q}_{+}}F_1^{A}(q^2)
+\frac{2M^2M_0^{\prime3}}{\tilde{Q}_{-}^2\tilde{Q}_{+}^2}F_2^{A}(q^2)
-\frac{4M^2M_{0}^{\prime2}(2M_0^2+M_0M_0^{\prime}+2(M_0^{\prime2}-q^2))}{\tilde{Q}_{-}^3\tilde{Q}_{+}^2}F_3^{A}(q^2)
+\frac{20M^2M_0^{\prime4}}{\tilde{Q}_{-}^3\tilde{Q}_{+}^2}F_4^{A}(q^2),\\
\end{split}
\end{equation}
where
\begin{equation}
\begin{split}
F_{(1,2,3,4)}^V(q^2)=&\int\bigg{(}\frac{dx_1d^2\vec{k}_{1\bot}}{2(2\pi)^3}\bigg{)}\bigg{(}\frac{dx_2d^2\vec{k}_{2\bot}}{2(2\pi)^3}\bigg{)}
\frac{\psi_b(x_i,\vec{k}_{i\bot})\psi_c^{\ast}(x_i^{\prime},\vec{k}^{\prime}_{i\bot})}{\sqrt{x_3x_3^{\prime}}}A_0B_0^{\prime}\text{Tr}[\cdots]\\
&\times\text{Tr}\big{[}(G_{\mathcal{B}_c})_{\beta\alpha}K^{\prime\alpha}(\slashed{p}_{3}^{\prime}+m_{3}^{\prime})\gamma^{\mu}\gamma^{5}(\slashed{p}_{3}+m_{3})(\slashed{P}+M_0)\Gamma_{(1,2,3,4),\mu}^{A,\beta}\big{]}.
\label{eq:formfactorsA}
\end{split}
\end{equation}
\end{widetext}

All the traces in Eq.~\eqref{eq:formfactorsV}, Eq.~\eqref{eq:trace}, and Eq.~\eqref{eq:formfactorsA} are calculable with the help of the FEYNCALC program~\cite{Mertig:1990an,Shtabovenko:2016sxi,Shtabovenko:2020gxv}, where the following relations
\begin{equation}
\begin{split}
P\cdot P&=M_{0}^{2},~~~~P^{\prime}\cdot P^{\prime}=M_{0}^{\prime2},\\
P\cdot P^{\prime}&=(M_0^2+M_0^{\prime2}-q^2)/2,\\
p_1\cdot P&=e_1M_0,~~~~p_2\cdot P=e_2M_0,\\
p_1\cdot P^{\prime}&=e_1^{\prime}M_0^{\prime},~~~~p_2\cdot P^{\prime}=e_2^{\prime}M_0^{\prime},\\
p_1\cdot p_2&=(M_0^2+m_3^2-m_1^2-m_2^2-2e_3M_0)/2,
\end{split}
\end{equation}
are used. We also have $p_{i}^{(\prime)2}=m_{i}^{(\prime)2}$ with $m_{i}^{(\prime)}$ being the mass of corresponding quark. Moreover, $e_i^{(\prime)}$, the energy of $i^{(\prime)}$-th quark, is defined as
\begin{equation}
e_{i}^{\prime}=\frac{1}{2}\Big{(}x_i^{(\prime)}M_0^{(\prime)}+\frac{m_{i}^{(\prime)2}+\vec{k}_{i\bot}^{(\prime)2}}{x_{i}^{(\prime)}M_{0}^{(\prime)}}\Big{)}.
\end{equation}

\section{The semirelativistic potential model for getting baryon wave function}
\label{sec3}

In Refs.~\cite{Chua:2018lfa,Ke:2019smy,Chua:2019yqh,Ke:2019lcf}, the spatial wave function of baryon is usually treated as a simple harmonic oscillator form with a phenomenological parameter $\beta$, which results in  the $\beta$ dependence of the calculated form factors.
For avoiding the  $\beta$ dependence of result,
we can take the numerical spatial wave function as input, which is obtained by solving the three-body Schr\"{o}dinger equation with the semirelativistic potential model. So, in the present section, we should introduce the semirelativistic potential model and the GEM.

In the study of baryon spectroscopy, the baryon wave function and its mass can be obtained by solving the Schr\"{o}dinger equation
\begin{equation}
\mathcal{H}\vert\Psi_{\mathbf{J},\mathbf{M_J}}\rangle=E\vert\Psi_{\mathbf{J},\mathbf{M_J}}\rangle
\label{eq:SchrodingerEquation}
\end{equation}
with the Rayleigh-Ritz variational principle, where $\mathcal{H}$ is the Hamiltonian and $E$ is the corresponding eigenvalue. In this calculation, the semirelativistic potential, which was given in Ref.~\cite{Capstick:1985xss}, are applied. The concerned Hamiltonian~\cite{Capstick:1985xss,Li:2021qod,Li:2021kfb}
\begin{equation}
\mathcal{H}=K+\sum_{i<j}(S_{ij}+G_{ij}+V^{\text{so(s)}}_{ij}+V^{\text{so(v)}}_{ij}+V^{\text{ten}}_{ij}+V^{\text{con}}_{ij})
\end{equation}
includes the kinetic energy $K=\sum_{i=1,2,3}\sqrt{m_i^2+p_i^2}$, the linear confinement term $S_{ij}$:
\begin{eqnarray}
S_{ij}&=&-\frac{3}{4}\left(br_{ij}\left[\frac{e^{-\sigma_{ij}^2r_{ij}^2}}{\sqrt{\pi}\sigma_{ij}r_{ij}}
+\left(1+\frac{1}{2\sigma_{ij}^2r_{ij}^2}\right)\frac{2}{\sqrt{\pi}}\right.\right.\nonumber\\
&&\left.\left.\times\int_{0}^{\sigma_{ij}r_{ij}}e^{-x^2}dx\right]\right)\mathbf{F_i}\cdot\mathbf{F_j}+\frac{c}{3}
\end{eqnarray}
with
\begin{equation}
\sigma_{ij}^2=\sigma_0^2\left[\frac{1}{2}+\frac{1}{2}\bigg{(}\frac{4m_im_j}{(m_i+m_j)^2}\bigg{)}^4+s^2\bigg{(}\frac{2m_im_j}{m_i+m_j}\bigg{)}^2\right],
\end{equation}
the Coulomb-like potential $G_{ij}$:
\begin{equation}
G_{ij}=\sum_{k}\frac{\alpha_k}{r_{ij}}\left[\frac{2}{\sqrt{\pi}}\int_{0}^{\tau_k r_{ij}}e^{-x^2}dx\right]\mathbf{F_i}\cdot\mathbf{F_j},
\label{eq:potentialG}
\end{equation}
the scalar typed spin-orbit interaction $V^{\text{so}(s)}$:
\begin{equation}
V^{\text{so}(s)}_{ij}=-\frac{\mathbf{r_{ij}}\times\mathbf{p_{i}}\cdot\mathbf{S_i}}{2m_i^2}\frac{1}{r_{ij}}
\frac{\partial S_{ij}}{\partial r_{ij}}+\frac{\mathbf{r_{ij}}\times\mathbf{p_{j}}\cdot\mathbf{S_j}}{2m_j^2}\frac{1}{r_{ij}}\frac{\partial S_{ij}}{\partial r_{ij}},
\end{equation}
the vector typed spin-orbit interaction $V^{\text{so}(v)}$:
\begin{equation}
\begin{split}
V^{\text{so}(v)}_{ij}=&\frac{\mathbf{r_{ij}}\times\mathbf{p_{i}}\cdot\mathbf{S_i}}{2m_i^2}\frac{1}{r_{ij}}\frac{\partial G_{ij}}{\partial r_{ij}}
-\frac{\mathbf{r_{ij}}\times\mathbf{p_{j}}\cdot\mathbf{S_j}}{2m_j^2}\frac{1}{r_{ij}}\frac{\partial G_{ij}}{\partial r_{ij}}\\
&-\frac{\mathbf{r_{ij}}\times\mathbf{p_{j}}\cdot\mathbf{S_i}-\mathbf{r_{ij}}\times\mathbf{p_{i}}\cdot\mathbf{S_j}}{m_i~m_j}
\frac{1}{r_{ij}}\frac{\partial G_{ij}}{\partial r_{ij}},
\end{split}
\end{equation}
the tensor potential $V^{\text{tens}}$:
\begin{equation}
\begin{split}
V^{\text{tens}}_{ij}=&-\frac{1}{m_im_j}\left[\left(\mathbf{S_i}\cdot\mathbf{\hat r_{ij}}\right)\left(\mathbf{S_j}\cdot \mathbf{\hat r_{ij}}\right)-\frac{\mathbf{S_i}\cdot\mathbf{S_j}}{3}\right]\\
&\times\left(\frac{\partial^2G_{ij}}{\partial r_{ij}^2}-\frac{\partial G_{ij}}{r_{ij}\partial r_{ij}}\right),
\end{split}
\end{equation}
and the spin-dependent contact potential $V^{\text{con}}$:
\begin{equation}
V^{\text{con}}_{ij}=\frac{2\mathbf{S_i}\cdot\mathbf{S_j}}{3m_i m_j}\nabla^2G_{ij},
\end{equation}
where $m_i$ stands for the mass of constituent quark $i$, and the $\mathbf{S_i}$ is the corresponding spin operator. $\langle\mathbf{F_i}\cdot\mathbf{F_j}\rangle=-2/3$ is for quark-quark interaction~\cite{Godfrey:1985xj}.
It is worthy to note that the running coupling constant $\alpha_{s}$ is defined as~\cite{Godfrey:1985xj,Capstick:1985xss}
\begin{equation}
\alpha_{s}(r)=\sum_{k=1}^{3}\alpha_{k}\frac{2}{\sqrt{\pi}}\int_{0}^{\gamma_{k}r}e^{-x^{2}}dx
\end{equation}
in Eq.~\eqref{eq:potentialG}. Here, $\{\alpha_1,\alpha_2,\alpha_3\}=\{0.25,0.15,0.20\}$, and
\begin{equation}
\tau_{k}=\frac{\gamma_{k}\sigma_{ij}}{\sqrt{\gamma_{k}^{2}+\sigma_{ij}^{2}}}
\end{equation}
with $\{\gamma_1,\gamma_2,\gamma_3\}=\{1/2,\sqrt{10}/2,\sqrt{1000}/2\}$. The remaining parameters are collected into Table~\ref{tab:parametersofGI}.

For partially compensating relativistic effect in the non-relativistic limit, the following transformation~\cite{Godfrey:1985xj,Capstick:1985xss}
\begin{equation}
\begin{split}
&G_{ij}\to\left(1+\frac{p^2}{E_iE_j}\right)^{1/2} G_{ij}\left(1+\frac{p^2}{E_iE_j}\right)^{1/2},\\
&\frac{V^{k}_{ij}}{m_im_j}\to\left(\frac{m_im_j}{E_iE_j}\right)^{1/2+\epsilon_k}\frac{V^k_{ij}}{m_im_j}\left(\frac{m_im_j}{E_iE_j}\right)^{1/2+\epsilon_k}
\end{split}
\end{equation}
should be made, where $E_i=\sqrt{p^2+m_i^2}$ is the energy of $i$-th constituent quark, the subscript $k$ are used to distinguish the contributions from the contact, tensor, vector spin-orbit, and scalar spin-orbit terms, and the $\epsilon_k$ are used to represent the relevant modification parameters.

By fitting the mass spectrum of the single charmed and single bottom baryon, the model parameters in the semirelativistic potential model are obtained as collected in Table~\ref{tab:parametersofGI}.

\begin{table}[htbp]\centering
\caption{The parameters adopted in the semirelativistic potential model~\cite{Li:2022nim}. Besides, the quark masses are chosen as $m_{u}=220\ \text{MeV}$, $m_{d}=220\ \text{MeV}$, $m_{s}=419\ \text{MeV}$, $m_{c}=1628\ \text{MeV}$, and $m_{b}=4977\ \text{MeV}$~\cite{Godfrey:1985xj,Capstick:1985xss}.}
\label{tab:parametersofGI}
\renewcommand\arraystretch{1.05}
\begin{tabular*}{80mm}{c@{\extracolsep{\fill}}ccc}
\toprule[1pt]
\toprule[0.5pt]
Parameters     & Values   & Parameters   & Values\\
\toprule[0.5pt]
$b~(\text{GeV}^2)$        &$0.1466\pm0.0007$      &$\epsilon^{\text{so}(s)}$   &$0.5000\pm0.0762$\\
$c~(\text{GeV})  $        &$-0.3490\pm0.0050$     &$\epsilon^{\text{so}(v)}$   &$-0.1637\pm0.0131$\\
$\sigma_0~(\text{GeV})$   &$1.7197\pm0.0304$      &$\epsilon^{\text{tens}}$    &$-0.3790\pm0.5011$\\
$s$                       &$0.5278\pm0.0718$      &$\epsilon^{\text{con}}$     &$-0.1612\pm0.0015$\\
\bottomrule[0.5pt]
\bottomrule[1pt]
\end{tabular*}
\end{table}

\begin{figure}[htbp]\centering
  \includegraphics[width=6cm]{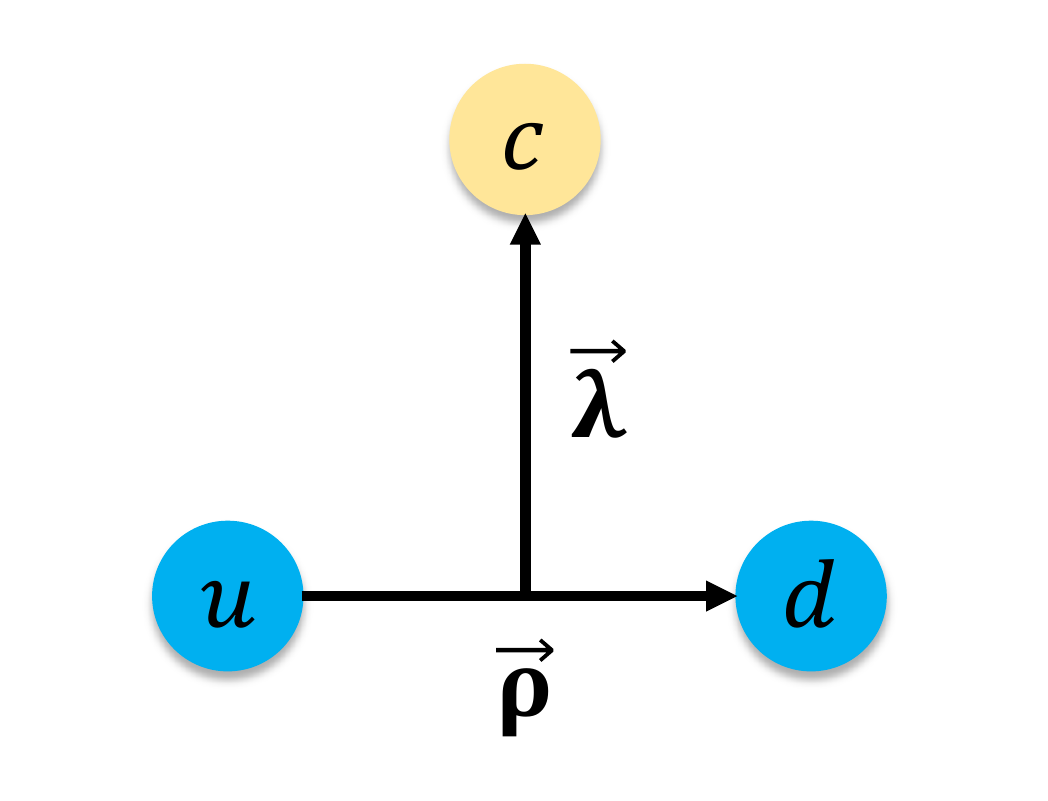}\\
  \caption{The definitions of internal Jacobi coordinates $\vec{\rho}$ and $\vec{\lambda}$ when taking the $\Lambda_c$ baryon as an example.}
  \label{fig:Jacobi}
\end{figure}

The total wave function of a baryon can be written as
\begin{equation}
\begin{split}
\Psi_{\mathbf{J},\mathbf{M_J}}=&\sum_{\alpha}C^{(\alpha)}\Psi_{\mathbf{J},\mathbf{M_J}}^{(\alpha)},\\
\Psi_{\mathbf{J},\mathbf{M_J}}^{(\alpha)}=&\chi^{\text{color}}
\left\{{\chi^\text{spin}}_{\mathbf{S},\mathbf{M_S}}
\psi^{\text{partial}}_{\mathbf{L},\mathbf{M_L}}\right\}_{\mathbf{J},\mathbf{M_J}}
\psi^{\text{flavor}},
\label{eq:wavefunction}
\end{split}
\end{equation}
which is composed of color, spin, spatial, and flavor terms, where $C^{(\alpha)}$ denotes the coefficient with $\alpha$ being the possible quantum number. The color wave function $\chi^{\text{color}}=(rgb-rbg+gbr-grb+brg-bgr)/\sqrt{6}$ is universal for any baryons. In the SU(2) flavor symmetry, the flavor wave function is expressed as $\psi_{\Lambda_{Q}}^{\text{flavor}}=(ud-du)Q/\sqrt{2}$ for the $\Lambda_{Q}$-typed baryon, while the flavor wave function is $\psi_{\Xi_{Q}}^{\text{flavor}}=(ns-sn)Q/\sqrt{2}$ with $n=u\ (\text{or}\ d)$ and $Q=b\ (\text{or}\ c)$ for a $\Xi_{Q}$-typed baryon. The subscripts \textbf{S} and \textbf{L} represent the total spin and total orbital angular momentum, respectively. And $\psi^{\text{spatial}}_{\mathbf{L},\mathbf{M_L}}$ is the spatial wave function of $\rho$-mode and $\lambda$-mode excitation
\begin{equation}
\psi^{\text{spatial}}_{\mathbf{L},\mathbf{M_L}}=
\left\{\phi_{\pmb{l_{\rho}},\pmb{ml_{\rho}}}
\phi_{\pmb{l_{\lambda}},\pmb{ml_{\lambda}}}\right\}_{\mathbf{L},\mathbf{M_L}},
\end{equation}
where the subscripts $\pmb{l_{\rho}}$ and $\pmb{l_{\lambda}}$ are the orbital angular momentum for the $\rho$ and $\lambda$-mode excitation, respectively. The single heavy baryon can be regarded as a bound state of light quark cluster and heavy quark.
Here, the $\rho$-mode indicates the radial excitation between two light quarks, while the $\lambda$-mode stands for the redial excitation between the light quark cluster and heavy quark. For the concerned bottom and charmed baryons, the internal Jacobi coordinates can be chosen as
\begin{equation}
\vec{\rho}=\vec{r}_2-\vec{r}_1,~~~\vec{\lambda}=\vec{r}_3-\frac{m_1\vec{r}_1+m_2\vec{r}_2}{m_1+m_2}.
\end{equation}
{For easily illustrating this point, we take the $\Lambda_c$ resonance as an example and present the definitions of the $\rho$-mode and $\lambda$-mode as displayed in Fig.~\ref{fig:Jacobi}.}

In the realistic calculation, the Gaussian basis~\cite{Hiyama:2003cu,Hiyama:2012sma,Yoshida:2015tia}
\begin{equation}
\begin{split}
\phi_{nlm}^{G}(\vec{r})=&\phi^{G}_{nl}(r)~Y_{lm}(\hat{r})\\
=&\sqrt{\frac{2^{l+2}(2\nu_{n})^{l+3/2}}{\sqrt{\pi}(2l+1)!!}}\lim_{\varepsilon\rightarrow0}
\frac{1}{(\nu_{n}\varepsilon)^l}\sum_{k=1}^{k_{\text{max}}}C_{lm,k}e^{-\nu_{n}(\vec{r}-\varepsilon\vec{D}_{lm,k})^2}
\label{eq:Gaussianbasis}
\end{split}
\end{equation}
is adopted to expand the spatial wave functions $\phi_{\pmb{l_{\rho}},\pmb{ml_{\rho}}}$ and $\phi_{\pmb{l_{\lambda}},\pmb{ml_{\lambda}}}$ ($n=1,2,\cdots,n_{max}$), where the Gaussian size parameter $\nu_{n}$ can be settled as a geometric progression~\cite{Luo:2022cun}
\begin{equation}
\nu_{n}=1/r^2_{n}, ~~~r_{n}=r_{min}~a^{n-1}
\end{equation}
with
\begin{equation}
a=\left(\frac{r_{max}}{r_{min}}\right)^{\frac{1}{n_{max}-1}}.
\end{equation}
The Gaussian basis in the momentum space $\phi_{nlm}^{G}(\vec{k})$ can be obtained by the replacement $\vec{r}\to\vec{k}$ and $\nu_{n}\to1/(4\nu_{n})$ in Eq.~\eqref{eq:Gaussianbasis}. In our calculation, the values of $\rho_{min}$ and $\rho_{max}$ are set as $0.2$~fm and $2.0$~fm, respectively, and $n_{\rho_{max}}=6$. In the meantime, the same Gaussian sized parameters are also applied to the $\lambda$-mode excitation.

With above preparation, we can calculate the kinematic, the potential, and the normalize matrix elements as
\begin{equation}
\begin{split}
T^{\alpha^{\prime},\alpha}=&\langle\Psi_{\mathbf{J},\mathbf{M_J}}^{(\alpha^{\prime})}\vert K \vert\Psi_{\mathbf{J},\mathbf{M_J}}^{(\alpha)}\rangle,\\
V^{\alpha^{\prime},\alpha}=&\langle\Psi_{\mathbf{J},\mathbf{M_J}}^{(\alpha^{\prime})}\vert V \vert\Psi_{\mathbf{J},\mathbf{M_J}}^{(\alpha)}\rangle,\\
N^{\alpha^{\prime},\alpha}=&\langle\Psi_{\mathbf{J},\mathbf{M_J}}^{(\alpha^{\prime})}\vert\Psi_{\mathbf{J},\mathbf{M_J}}^{(\alpha)}\rangle.
\label{eq:matrixelement}
\end{split}
\end{equation}
Then, the Schr\"{o}dinger equation in Eq.~\eqref{eq:SchrodingerEquation} can be solved by the Rayleigh-Ritz variational principle as
\begin{equation}
\Big{(}T^{\alpha^{\prime},\alpha}+V^{\alpha^{\prime},\alpha}\Big{)}C^{(\alpha)}=EN^{\alpha^{\prime},\alpha}C^{(\alpha)}.
\end{equation}

{For clarity, we take the $\Lambda_{c}(2625)$ as an example to illustrate the detailed matrix element defined in Eq.~\eqref{eq:matrixelement}. The quantum numbers of $\Lambda_{c}(2625)$ are $(\alpha)=(l_{\rho},l_{\lambda},L,S_{\rho},S,J)=(0,1,1,0,1/2,3/2)$. By expanding the wave function in Eq.~\eqref{eq:wavefunction} with $n_{\rho_{\text{max}}}\times n_{\lambda_{\text{max}}}=6\times6=36$ Gaussian bases in Eq.~\eqref{eq:Gaussianbasis},  the matrix element $T^{\alpha^{\prime},\alpha}$ can be written as
\begin{equation}
T^{\alpha^{\prime},\alpha}=
\begin{bmatrix}
T_{1,1,1,1}   &\cdots   &\cdots   &\cdots\\
\vdots &\ddots   &         &\\
\vdots &         &T_{n_{\rho}^{\prime},n_{\lambda}^{\prime},n_{\rho},n_{\lambda}}     &\\
\vdots &         &         &\ddots\\
\end{bmatrix}_{36\times36},
\end{equation}
where
\begin{equation}
\begin{split}
T_{n_{\rho}^{\prime},n_{\lambda}^{\prime},n_{\rho},n_{\lambda}}=&
\langle \chi^\text{spin}_{S^{\prime},M_{S}^{\prime}}\Big{\{}\phi_{n_{\rho}^{\prime}l_{\rho}^{\prime}m_{l_{\rho}}^{\prime}}^{G}(\vec{p}_{\rho})
\phi_{n_{\lambda}^{\prime}l_{\lambda}^{\prime}m_{l_{\lambda}}^{\prime}}^{G}(\vec{p}_{\lambda})\Big{\}}_{L^{\prime},M_{L}^{\prime}}
\vert K \vert
\chi^\text{spin}_{S,M_{S}}\\
&\times\Big{\{}\phi_{n_{\rho}l_{\rho}m_{l_{\rho}}}^{G}(\vec{p}_{\rho})
\phi_{n_{\lambda}l_{\lambda}m_{l_{\lambda}}}^{G}(\vec{p}_{\lambda})\Big{\}}_{L,M_{L}}
\rangle.
\end{split}
\end{equation}
Here, we neglect the contributions of the color and flavor wave functions, since their overlap equals to 1. The matrix elements $V^{\alpha^{\prime},\alpha}$ and $N^{\alpha^{\prime},\alpha}$ can also be obtained in similar method.}

Now, we can handle the Schr\"{o}dinger equation to obtain the eigenvectors and eigenvalues, which correspond to the baryon wave functions and the masses, respectively. In Table~\ref{tab:wavefunctions}, we present our results of the masses and the radial components of the spatial wave functions of the concerned baryons. It is obvious that the calculated masses are consistent with the experimental values~\cite{ParticleDataGroup:2022pth}. It also indicates that we can well reproduce the charmed and bottom baryon spectrum by the adopted potential model, and the obtained numerical wave functions are as input  when getting the form factors of these discussed weak transitions.

\begin{table*}[htbp]\centering
\caption{The comparisons of the masses by our calculations and the PDG values~\cite{ParticleDataGroup:2022pth}, and the radial components of spatial wave functions of the concerned bottomed baryons $\Lambda_b $ and $\Xi_b$, as well as the $P$-wave charmed baryons $\Lambda_c(2625)$ and $\Xi_c(2815)$ from the GI model and GEM. The Gaussian bases $(n_{\rho},n_{\lambda})$ listed in the forth column are arranged as $[(1,1),(1,2),\cdots,(1,n_{\lambda_{max}}),(2,1),(2,2), \cdots,(2,n_{\lambda_{max}}),\cdots,(n_{\rho_{max}},1), (n_{\rho_{max}},2),\cdots,(n_{\rho_{max}},n_{\lambda_{max}})]$. For the masses of the $\Xi_b$ and $\Xi_c(2815)$, the values for the neutral and charged states are degenerated in our calculation since the same mass for the $u$ and $d$ quarks is applied in the potential model.}
\label{tab:wavefunctions}
\renewcommand\arraystretch{1.05}
\begin{tabular*}{172mm}{c@{\extracolsep{\fill}}cccc}
\toprule[1pt]
\toprule[0.5pt]
States  &This work (GeV)  &Experiment (MeV)~\cite{ParticleDataGroup:2022pth}  &Eigenvectors\\
\midrule[0.5pt]
\multirow{9}*{\shortstack{$\Lambda_b^0$}}        &\multirow{9}*{$5.621\pm0.005$}    &\multirow{9}*{\makecell[c]{$5619.60\pm0.17$}}
&$\Big{[}0.0068\pm0.0007,0.0442\pm0.0014,0.0732\pm0.0016,0.0032\pm0.0003,$\\
&&&$0.0011\pm0.0001,-0.0004\pm0.0000,0.0270\pm0.0012,0.0204\pm0.0010,$\\
&&&$0.0273\pm0.0022,0.0067\pm0.0004,-0.0027\pm0.0001,0.0007\pm0.0000,$\\
&&&$-0.017\pm0.0002,0.2541\pm0.0058,0.2427\pm0.0006,0.0005\pm0.0002,$\\
&&&$0.0060\pm0.0001,-0.0017\pm0.0000,-0.0037\pm0.0003,-0.0426\pm0.0010,$\\
&&&$0.4052\pm0.0028,0.0253\pm0.0025,-0.0023\pm0.0007,0.0004\pm0.0002,$\\
&&&$0.0071\pm0.0001,-0.0052\pm0.0008,0.0105\pm0.0008,0.1224\pm0.0015,$\\
&&&$-0.0246\pm0.0001,0.0054\pm0.0000,-0.0020\pm0.0000,0.0010\pm0.0003,$\\
&&&$-0.0112\pm0.0003,-0.0139\pm0.0001,0.0086\pm0.0001,-0.0017\pm0.0000\Big{]}$\\
\multirow{9}*{\shortstack{$\Xi_b^{0,-}$}}        &\multirow{9}*{$5.809\pm0.004$}    &\multirow{9}*{\makecell[c]{$5791.9\pm0.5$\\ $5797.0\pm0.6$}}
&$\Big{[}0.0069\pm0.0008,0.0293\pm0.0012,0.0543\pm0.0016,-0.0002\pm0.0003,$\\
&&&$0.0014\pm0.0001,-0.0004\pm0.0000,0.0231\pm0.0013,0.0397\pm0.0003,$\\
&&&$0.0278\pm0.0018,0.0114\pm0.0003,-0.0037\pm0.0000,0.0009\pm0.0000,$\\
&&&$-0.0093\pm0.0003,0.2285\pm0.0053,0.2601\pm0.0007,-0.0165\pm0.0004,$\\
&&&$0.0100\pm0.0000,-0.0026\pm0.0000,-0.0043\pm0.0005,-0.0094\pm0.0001,$\\
&&&$0.3992\pm0.0037,0.0525\pm0.0026,-0.0092\pm0.0006,0.0019\pm0.0001,$\\
&&&$0.0048\pm0.0001,-0.0108\pm0.0005,0.0095\pm0.0005,0.0813\pm0.0015,$\\
&&&$-0.0145\pm0.0002,0.0033\pm0.0000,-0.0011\pm0.0000,0.0011\pm0.0002,$\\
&&&$-0.0052\pm0.0002,-0.0070\pm0.0001,0.0034\pm0.0001,-0.0007\pm0.0000\Big{]}$\\
\multirow{9}*{\shortstack{$\Lambda_c^+(2625)$}}        &\multirow{9}*{$2.623\pm0.007$}    &\multirow{9}*{\makecell[c]{$2628.11\pm0.19$}}
&$\Big{[}0.0012\pm0.0001,0.0148\pm0.0007,0.0760\pm0.0021,0.0359\pm0.0004,$\\
&&&$-0.0044\pm0.0001,0.0010\pm0.0000,0.0066\pm0.0003,0.0059\pm0.0002,$\\
&&&$0.0376\pm0.0018,0.0183\pm0.0015,-0.0034\pm0.0002,0.0008\pm0.0000,$\\
&&&$-0.0027\pm0.0000,0.0767\pm0.0022,0.2861\pm0.0039,0.1060\pm0.0001,$\\
&&&$-0.0126\pm0.0001,0.0027\pm0.0000,0.0031\pm0.0002,-0.0383\pm0.0002,$\\
&&&$0.2926\pm0.0013,0.2054\pm0.0037,-0.0346\pm0.0004,0.0082\pm0.0001,$\\
&&&$0.0028\pm0.0001,-0.0030\pm0.0001,-0.0008\pm0.0012,0.1395\pm0.0009,$\\
&&&$-0.0074\pm0.0003,0.0018\pm0.0001,-0.0010\pm0.0000,0.0017\pm0.0000,$\\
&&&$-0.0077\pm0.0004,-0.0222\pm0.0001,0.0072\pm0.0000,-0.0013\pm0.0000\Big{]}$\\
\multirow{9}*{\shortstack{$\Xi_c^{0,+}(2815)$}}        &\multirow{9}*{$2.811\pm0.006$}    &\multirow{9}*{\makecell[c]{$2819.79\pm0.30$\\ $2816.51\pm0.25$}}
&$\Big{[}0.0012\pm0.0001,0.0100\pm0.0005,0.0553\pm0.0020,0.0231\pm0.0005,$\\
&&&$-0.0029\pm0.0001,0.0006\pm0.0000,0.0065\pm0.0003,0.0119\pm0.0001,$\\
&&&$0.0432\pm0.0013,0.0252\pm0.0011,-0.0043\pm0.0001,0.0010\pm0.0000,$\\
&&&$-0.0017\pm0.0000,0.0715\pm0.0018,0.2859\pm0.0038,0.0892\pm0.0000,$\\
&&&$-0.0110\pm0.0000,0.0023\pm0.0000,0.0042\pm0.0001,-0.0307\pm0.0000,$\\
&&&$0.3138\pm0.0014,0.2328\pm0.0038,-0.0377\pm0.0004,0.0089\pm0.0001,$\\
&&&$0.0017\pm0.0000,-0.0036\pm0.0001,-0.0046\pm0.0008,0.1014\pm0.0011,$\\
&&&$-0.0049\pm0.0002,0.0012\pm0.0000,-0.0005\pm0.0000,0.0011\pm0.0000,$\\
&&&$-0.0032\pm0.0002,-0.0122\pm0.0000,0.0031\pm0.0000,-0.0006\pm0.0000\Big{]}$\\
\toprule[0.5pt]
\toprule[1pt]
\end{tabular*}
\end{table*}

\section{The form factors and weak decays}
\label{sec4}

\subsection{The weak transition form factors}

In the following, we calculate these involved form factors of the $\Lambda_b\to\Lambda_c(2625)$ and $\Xi_b\to\Xi_c(2815)$ transitions numerically. The masses of baryons are quoted from the Particle Data Group (PDG)~\cite{ParticleDataGroup:2022pth}, and the spatial wave functions illustrated in Sec.~\ref{sec3} are shown in Table~\ref{tab:wavefunctions}.

\begin{figure*}[htbp]\centering
  \begin{tabular}{cc}
  \includegraphics[width=70mm]{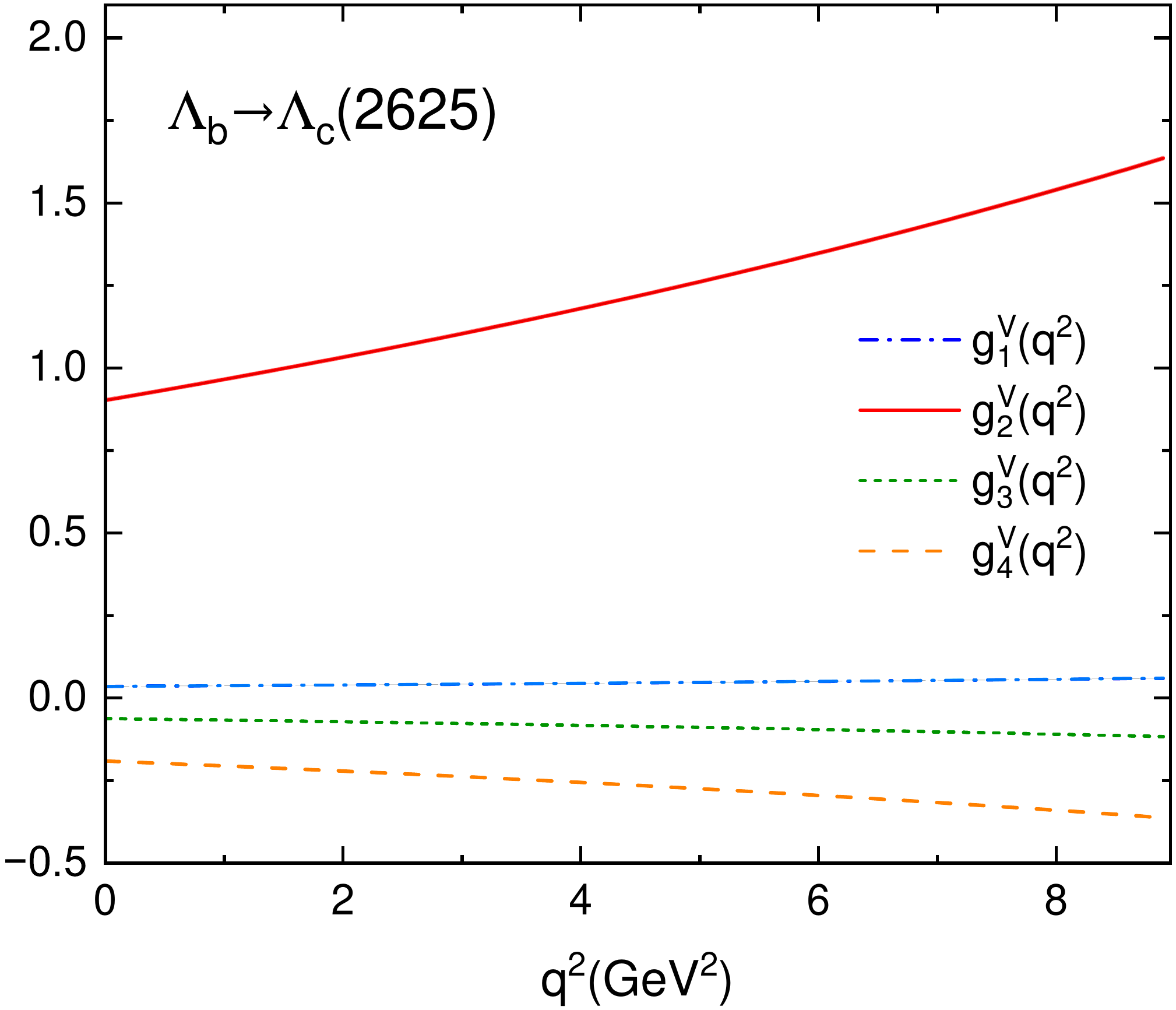}
  \includegraphics[width=70mm]{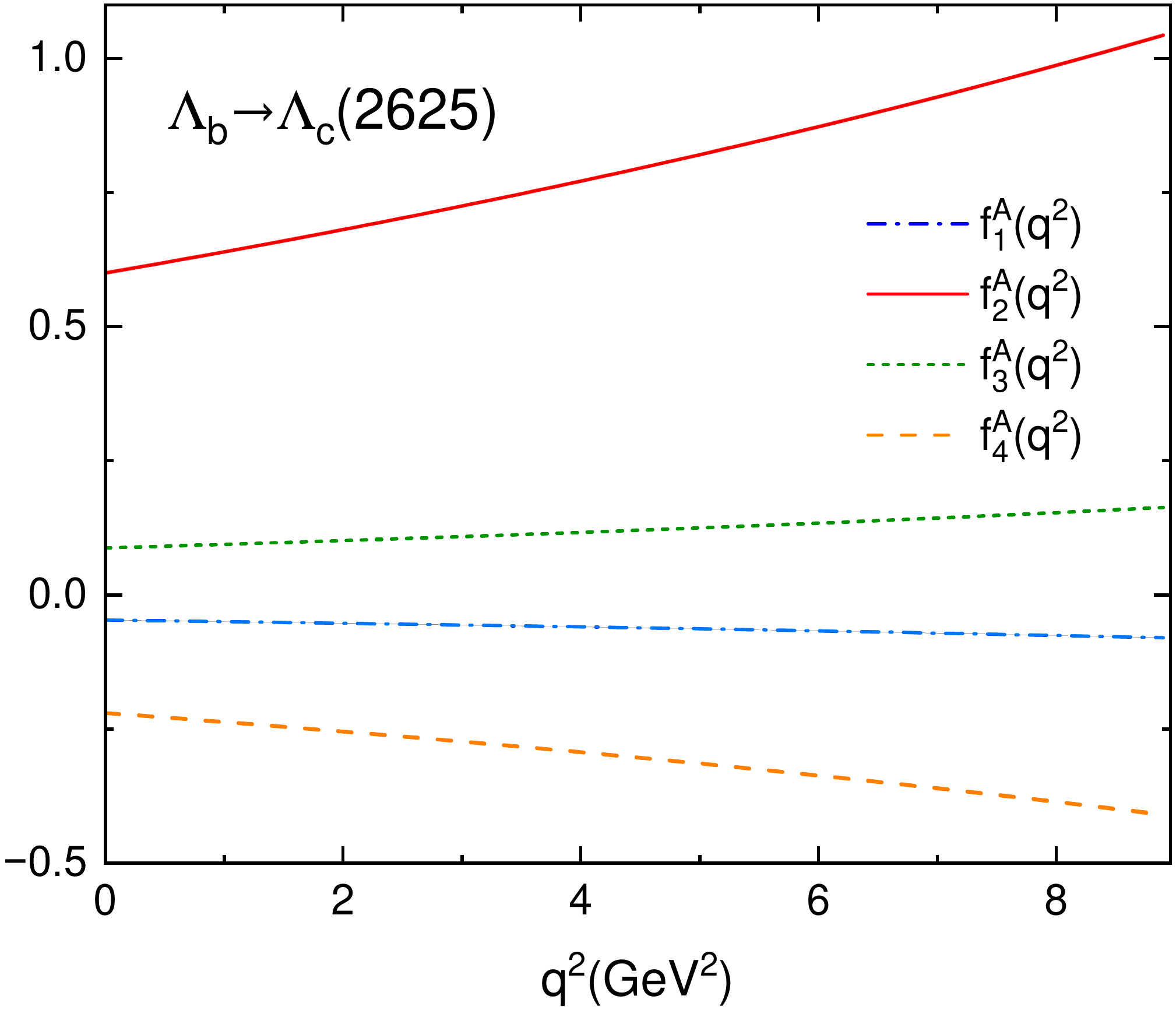}\\
  \includegraphics[width=70mm]{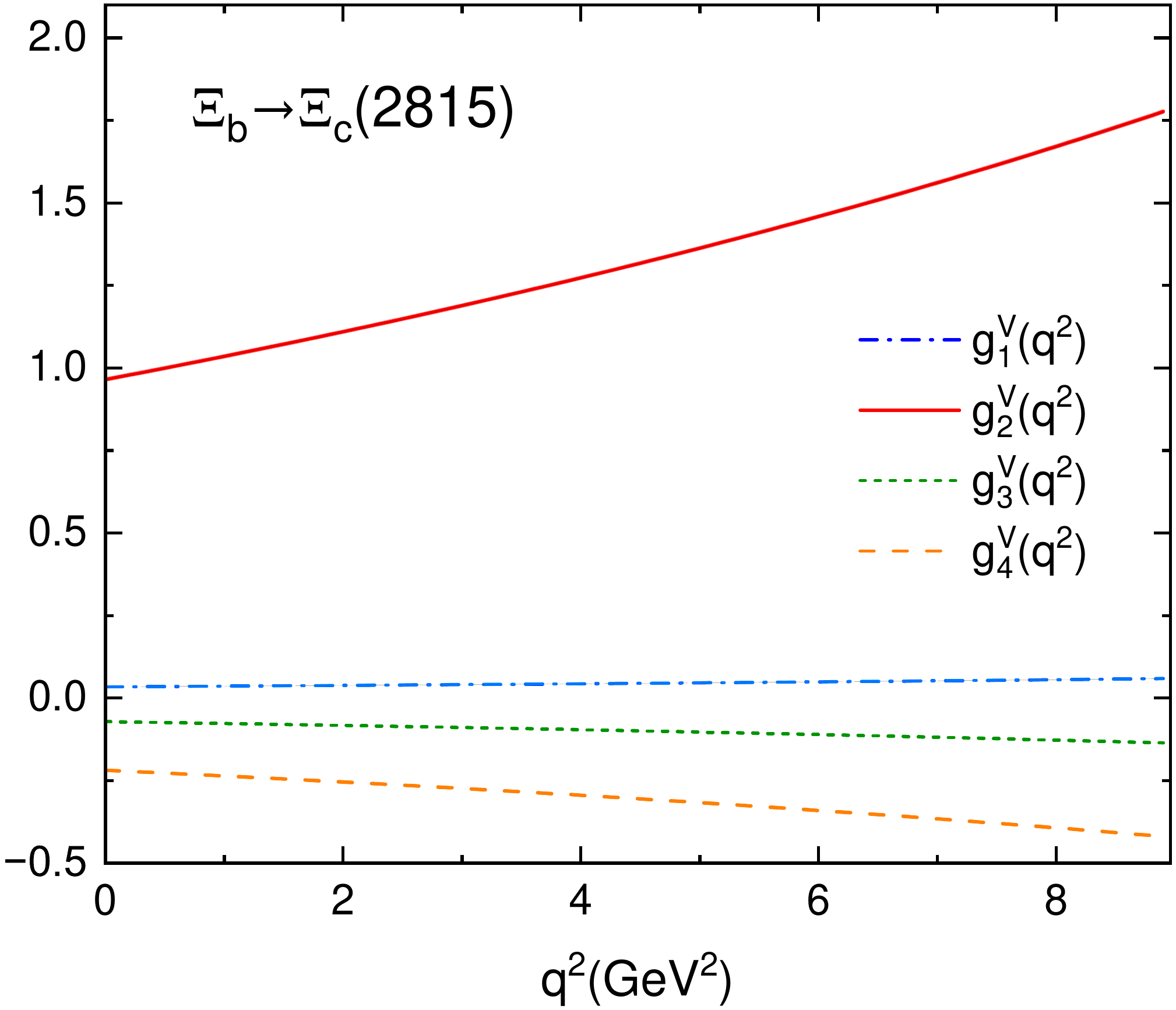}
  \includegraphics[width=70mm]{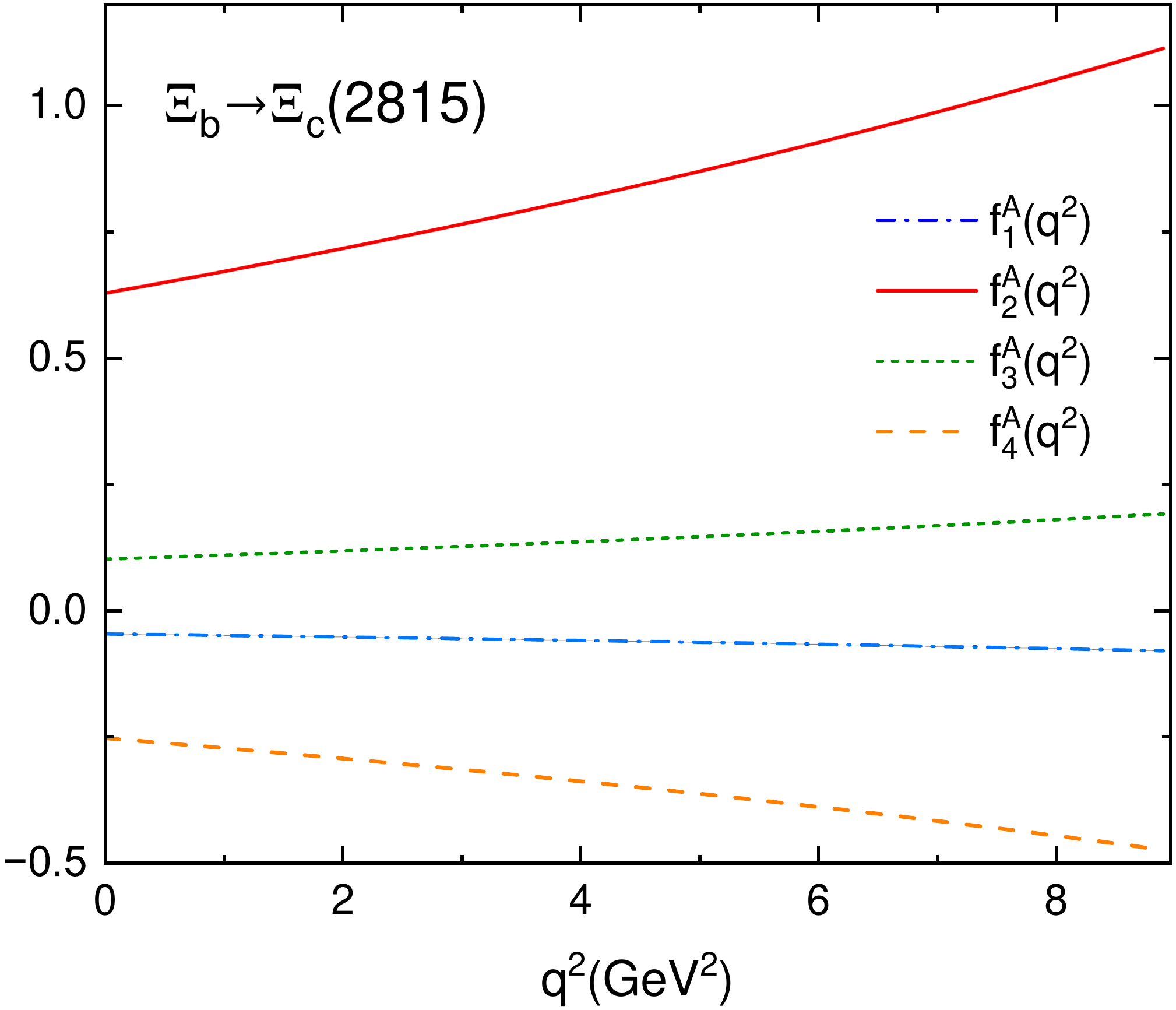}
  \end{tabular}
  \caption{The $q^2$ dependent form factors of the $\Lambda_b\to\Lambda_c(2625)$ (top panels) and $\Xi_b\to\Xi_c(2815)$ (bottom panels) transitions. Here, the uncertainties are also added. However, they are not obvious when we present the corresponding results.}
\label{fig:ffs}
\end{figure*}

Eq.~\eqref{eq:formfactorsV} and Eq.~\eqref{eq:formfactorsA} are worked in spacelike region ($q^2<0$), since we have set the $q^{+}=0$ condition. We need to extrapolate the obtained form factors to the timelike region ($q^2>0$). To do the extrapolation, we take advantage of the $z$-series parameterization as
\begin{equation}
f(q^2)=\frac{1}{1-q^2/(m_{\text{pole}}^f)^2}\Big{[}a_{0}+a_{1}z^{f}(q^2)\Big{]},
\label{eq:fitness}
\end{equation}
where $a_{0}^{f}$ and $a_{1}^{f}$ are free parameters needed to be fitted in spacelike region, and we have~\cite{Lellouch:1995yv,Bourrely:2005hp,Bourrely:2008za,Bharucha:2015bzk,Huang:2022lfr,Aliev:2022gxi}
\begin{equation}
z^f(q^2)=\frac{\sqrt{t_+^f-q^2}-\sqrt{t_+^f-t_0}}{\sqrt{t_+^f-q^2}+\sqrt{t_+^f-t_0}}
\end{equation}
with $t_{\pm}^f=(M \pm M^{\prime})^2$. The parameter $t_0$ is chosen as~\cite{Bharucha:2015bzk,Aliev:2022gxi}
\begin{equation}
0\leqslant t_0=t_{+}\bigg{(}1-\sqrt{1-\frac{t_{-}}{t_+}}\bigg{)}\leqslant t_{-}.
\end{equation}
The pole masses are chosen as $m_{B_c}=6.275\ \text{GeV}$~\cite{ParticleDataGroup:2022pth} for $g_{(1,3,4)}^V$, $m_{B_c^\ast}=6.338\ \text{GeV}$~\cite{Godfrey:2004ya} for $g_2^V$, $m_{B_{c0}}=6.706\ \text{GeV}$~\cite{Godfrey:2004ya} for $f_{(1,3,4)}^A$, and $m_{B_{c1}}=6.741\ \text{GeV}$~\cite{Godfrey:2004ya} for $f_2^A$.
In order to fix the free parameters $a_{0}^{f}$ and $a_{1}^{f}$, we numerically compute 24 points for each form factors from $q^2=-q_{\text{max}}^2$ to $q^2=-0.01\ \text{GeV}^2$ in the spacelike region, and then fit them with the MINUIT program. The fitted parameters are collected in Table~\ref{tab:fitness}, and the $q^2$ dependence of the form factors of $\Lambda_b\to\Lambda_c(2625)$ and $\Xi_b\to\Xi_c(2815)$ transitions are displayed in Fig.~\ref{fig:ffs}.

{In Table~\ref{tab:fitness}, we also present the $\chi^{2}$ values, which is defined by
\begin{equation}\chi^{2}=\frac{1}{n(n-1)}\sum_{i=1}^{n}\Bigg{(}\frac{f^{cal}(q_{i}^{2})-f^{ana}(q_{i}^{2})}{\delta f^{cal}(q_{i}^{2})}\Bigg{)}^{2},
\end{equation}
to characterize the analytical continuation, where $n=24$, the $f^{cal}$ and $f^{ana}$ represent the calculated value by quark model and the value of analytical continuation. The $\delta f^{cal}$ is the error of $f^{cal}$. Considering that the z-series parameterization have been widely used to  perform the analytical continuation, and the $\chi^{2}$ value in our fitting is suitable, in this work we take the $z$-series form to deal with the parameterization.}

\begin{table*}[htbp]\centering
\caption{The fitted parameters for the form factors of the $\Lambda_b\to\Lambda_c(2625)$ and $\Xi_b\to\Xi_c(2815)$ transitions in Eq.~\eqref{eq:fitness}.}
\label{tab:fitness}
\renewcommand\arraystretch{1.05}
\begin{tabular*}{156mm}{c@{\extracolsep{\fill}}ccccccc}
\toprule[1pt]
\toprule[0.5pt]
$f$         &$a_0$   &$a_1$   &$\chi^{2}$    &$f$   &$a_0$   &$a_1$   &$\chi^{2}$\\
\hline
\multicolumn{8}{c}{$\Lambda_b\to\Lambda_c(2625)$}  \\
\hline
$g_1^V$    &$0.0409\pm0.0002$   &$-0.3224\pm0.0053$   &$0.015$  &$f_1^A$    &$-0.0555\pm0.0003$    &$0.4855\pm0.0074$   &$0.026$\\
$g_2^V$    &$1.0889\pm0.0016$   &$-10.5808\pm0.0459$  &$0.460$  &$f_2^A$    &$0.7205\pm0.0005$     &$-6.8173\pm0.0168$  &$1.240$\\
$g_3^V$    &$-0.0763\pm0.0002$  &$0.8045\pm0.0046$    &$0.310$  &$f_3^A$    &$0.1093\pm0.0003$     &$-1.2327\pm0.0075$  &$0.363$\\
$g_4^V$    &$-0.2360\pm0.0004$  &$2.5599\pm0.0127$    &$0.473$  &$f_4^A$    &$-0.2749\pm0.0006$    &$3.1018\pm0.0169$   &$0.455$\\
\hline
\multicolumn{8}{c}{$\Xi_b\to\Xi_c(2815)$}  \\
\hline
$g_1^V$    &$0.0397\pm0.0002$   &$-0.3688\pm0.0065$   &$0.016$  &$f_1^A$    &$-0.0549\pm0.0003$    &$0.5577\pm0.0091$  &$0.028$\\
$g_2^V$    &$1.1750\pm0.0016$   &$-13.1039\pm0.0538$  &$0.564$  &$f_2^A$    &$0.7629\pm0.0006$     &$-8.3589\pm0.0203$  &$1.403$\\
$g_3^V$    &$-0.0884\pm0.0002$  &$1.0549\pm0.0064$    &$0.297$  &$f_3^A$    &$0.1281\pm0.0003$     &$-1.6232\pm0.0103$  &$0.361$\\
$g_4^V$    &$-0.2719\pm0.0004$  &$3.3205\pm0.0139$    &$0.710$  &$f_4^A$    &$-0.3167\pm0.0006$    &$3.9975\pm0.0189$  &$0.625$\\
\bottomrule[0.5pt]
\bottomrule[1pt]
\end{tabular*}
\end{table*}

In Table~\ref{tab:ffscomparison}, we compare our results of the form factors $g_{(1,2,3,4)}^{V}(q^2)$ and $f_{(1,2,3,4)}^{A}(q^2)$ at the $q^2=0$ and $q^2=q_{\text{max}}^2$ endpoints of the $\Lambda_b\to\Lambda_c(2625)$ transition with the other theoretical predictions by LFQM~\cite{Chua:2019yqh} and LQCD~\cite{Meinel:2021mdj}. The results of LQCD are reproduced with TABLE IX in Ref.~\cite{Meinel:2021mdj}.
In particular, the central value $O$ and the corresponding statistical uncertainty $\sigma_{O,\text{stat}}$ are reproduced by the so-called ``nominal-order" fitting, i.e.,
\begin{equation}
\begin{split}
f(q^2)&=F^{f}+A^{f}\big{(}\omega(q^2)-1\big{)},\\
\omega(q^2)&=\frac{M^2+M^{\prime2}-q^2}{2MM^{\prime}},
\end{split}
\end{equation}
and the systematic uncertainty can be obtained by
\begin{equation}
\sigma_{O,\text{syst}}=\text{max}\Big{(}\vert O_{\text{HO}}-O\vert,\sqrt{\vert \sigma_{O,\text{HO},\text{stat}}^2-\sigma_{O,\text{stat}}^2\vert}\Big{)},
\end{equation}
where $O_{\text{HO}}$ and $\sigma_{O,\text{HO},\text{stat}}$ are the central value and the corresponding statistical uncertainty in the ``higher-order" fitting:
\begin{equation}
f_{\text{HO}}(q^2)=F_{\text{HO}}^{f}+A_{\text{HO}}^{f}\big{(}\omega(q^2)-1\big{)}.
\end{equation}
Finally, the total uncertainty can be obtained by adding the systematic and statistical uncertainties in quadrature as
\begin{equation}
\sigma_{O,\text{total}}=\sqrt{\sigma_{O,\text{syst}}^2-\sigma_{O,\text{stat}}^2}.
\end{equation}

The definitions of the form factors used in LQCD~\cite{Meinel:2021rbm,Meinel:2021mdj} can be  converted into the present forms by the relations
in the Appendix 2 of Ref.~\cite{Meinel:2021rbm} combined with
\begin{equation}
\begin{split}
g_{1}^{V}=&F_{1}^{V},~~~g_{2}^{V}=F_{2}^{V},~~~g_{3}^{V}=F_{3}^{V}-M^{\prime}/M F_{4}^{V},~~~g_{4}^{V}=F_{4}^{V},\\
f_{1}^{A}=&F_{1}^{A},~~~f_{2}^{A}=F_{2}^{A},~~~f_{3}^{A}=F_{3}^{A}-M^{\prime}/M F_{4}^{A},~~~f_{4}^{A}=F_{4}^{A}.
\end{split}
\end{equation}
We emphasize that only the $q^{2}_{\text{max}}$ endpoint values are presented, since the LQCD's results are limited to small kinematic region near $q_{\text{max}}^{2}$. Our results of $g_{1,2}^{V}(q_{\text{max}}^{2})$ and $f_{1,2}^{A}(q_{\text{max}}^{2})$ are comparable with the LQCD's results, while others show some deviations. We expect more theoretical works on these form factors to further enrich our knowledge on these weak decays.

\begin{table*}[htbp]\centering
\caption{The theoretical predictions for the form factors $g_{(1,2,3,4)}^{V}(q^2)$ and $f_{(1,2,3,4)}^{A}(q^2)$ at $q^2=0$ and $q^2=q_{\text{max}}^2$ endpoints of the $\Lambda_b\to\Lambda_c(2625)$ transition using different approaches.}
\label{tab:ffscomparison}
\renewcommand\arraystretch{1.20}
\begin{tabular*}{156mm}{c@{\extracolsep{\fill}}cccc}
\toprule[1pt]
\toprule[0.5pt]
                           &$g_{1}^{V}(0)$                 &$g_{2}^{V}(0)$                  &$g_{3}^{V}(0)$
&$g_{4}^{V}(0)$\\
This Work                  &$0.0352\pm0.0002$              &$0.9023\pm0.0018$               &$-0.0621\pm0.0002$  &$-0.1909\pm0.0005$\\
LFQM~\cite{Chua:2019yqh}   &$-0.007^{+0.037}_{-0.026}$     &$0.509^{+0.184}_{-0.173}$       &$0.088^{+0.039}_{-0.043}$  &$0.004^{+0.058}_{-0.053}$\\
\hline
                           &$f_{1}^{A}(0)$                 &$f_{2}^{A}(0)$                  &$f_{3}^{A}(0)$
&$f_{4}^{A}(0)$\\
This Work                  &$-0.0469\pm0.0003$             &$0.6003\pm0.0006$               &$0.0876\pm0.0003$   &$-0.2202\pm0.0007$\\
LFQM~\cite{Chua:2019yqh}   &$0.028^{+0.065}_{-0.032}$      &$0.545^{+0.111}_{-0.104}$       &$0.022^{+0.033}_{-0.091}$  &$-0.005^{+0.104}_{-0.068}$\\
\hline
                           &$g_{1}^{V}(q_{\text{max}}^2)$   &$g_{2}^{V}(q_{\text{max}}^2)$
&$g_{3}^{V}(q_{\text{max}}^2)$   &$g_{4}^{V}(q_{\text{max}}^2)$\\
This Work                  &$0.0603\pm0.0003$              &$1.6412\pm0.0023$               &$-0.1171\pm0.0003$  &$-0.3639\pm0.0006$\\
LFQM~\cite{Chua:2019yqh}   &$-0.009^{+0.046}_{-0.033}$     &$0.737^{+0.267}_{-0.251}$       &$0.115^{+0.051}_{-0.056}$  &$0.005^{+0.072}_{-0.066}$\\
LQCD~\cite{Meinel:2021mdj}       &$0.0692\pm0.0045$     &$1.1340\pm0.1556$     &$-0.7977\pm0.3646$    &$0.2117\pm0.2795$\\
\hline
                           &$f_{1}^{A}(q_{\text{max}}^2)$   &$f_{2}^{A}(q_{\text{max}}^2)$
&$f_{3}^{A}(q_{\text{max}}^2)$    &$f_{4}^{A}(q_{\text{max}}^2)$\\
This Work                  &$-0.0800\pm0.0004$              &$1.0470\pm0.0007$               &$0.1636\pm0.0004$   &$-0.4115\pm0.0008$\\
LFQM~\cite{Chua:2019yqh}   &$0.035^{+0.082}_{-0.040}$  &$0.756^{+0.154}_{-0.144}$  &$0.027^{+0.041}_{-0.114}$  &$-0.006^{+0.131}_{-0.086}$\\
LQCD~\cite{Meinel:2021mdj}       &$-0.0660\pm0.0280$     &$0.8310\pm0.0978$     &$1.3386\pm3.4803$    &$0.1795\pm2.9081$\\
\bottomrule[0.5pt]
\bottomrule[1pt]
\end{tabular*}
\end{table*}

Besides, by Heavy Quark Effective Theory (HQET), one can rewrite the weak transition matrix element of the concerned $\mathcal{B}_{b}(\bar{3}_{f},1/2^{+})\to\mathcal{B}_{c}(\bar{3}_{f},3/2^{-})$ transition as~\cite{Chua:2019yqh}
\begin{equation}
\langle \mathcal{B}_{c}(v^{\prime})\vert j_{V-A}^{\mu} \vert\mathcal{B}_{b}(v)\rangle=
-\sigma(\omega)\bar{u}_{\alpha}(v^{\prime})v^{\alpha}\gamma^{\mu}(1-\gamma^{5})u(v),
\end{equation}
where $v=P/M$ and $v^{\prime}=P^{\prime}/M^{\prime}$ are the velocities of the initial and the final baryons, respectively. Thus, the form factors have simpler behavior~\cite{Chua:2019yqh}
\begin{equation}
g_{2}^{V}=f_{2}^{A}=\sigma(\omega),~~~g_{1,2,3}^{V}=f_{1,2,3}^{A}=0.
\end{equation}
As shown in Fig.~\ref{fig:ffs}, obviously our results of $g_{2}^{V}$ and $f_{2}^{A}$ are apparently larger than those of $g_{(1,3,4)}^{V}$ and $f_{(1,3,4)}^{A}$, which is consistent with the expectation from HQET.

\subsection{The semileptonic decays}

In this section, we further calculate the semileptonic decays $\Lambda_b^{0}\to\Lambda_c^{+}(2625)\ell^{-}\nu_{\ell}$ and $\Xi_b^{0,-}\to\Xi_c(2815)^{+,0}\ell^{-}\nu_{\ell}$ $(\ell^{-}=e^{-},\mu^{-},\tau^{-})$. The differential decay width of the semileptonic decay can be obtained by
\begin{equation}
\begin{split}
\frac{d^{2}\Gamma}{dq^2d\cos\theta_{\ell}}=&\Big{\vert}\frac{G_{F}}{\sqrt{2}}V_{cb}\Big{\vert}^{2}
\frac{\sqrt{Q_{+}Q_{-}}q^{2}(1-\hat{m_{\ell}}^{2})^{2}}{512\pi^{3}M^{3}}\\
&\times\Big{(}L_{1}+L_{2}\cos\theta_{\ell}+L_{3}\cos2\theta_{\ell}\Big{)},
\end{split}
\end{equation}
where $Q_{\pm}=(M\pm M^{\prime})^2-q^2$, $\hat{m_{\ell}}^{2}=m_{\ell}^2/q^2$ and the angular coefficients $L_{1}$, $L_{2}$, and $L_{3}$ are given as
\begin{equation}
\begin{split}
L_{1}=&\frac{1}{2}(3+\hat{m_{\ell}}^{2})
\Big{(}H_{-3/2,-1}^{2}+H_{-1/2,-1}^{2}+H_{+1/2,+1}^{2}+H_{+3/2,+1}^{2}\Big{)}\\
&+(1+\hat{m_{\ell}}^{2})\Big{(}H_{+1/2,0}^{2}+H_{-1/2,0}^{2}\Big{)}\\
&+2\hat{m_{\ell}}^{2}\Big{(}H_{+1/2,t}^{2}+H_{-1/2,t}^{2}\Big{)},
\end{split}
\end{equation}
\begin{equation}
\begin{split}
L_{2}=&2\Big{(}H_{-3/2,-1}^{2}+H_{-1/2,-1}^{2}-H_{+1/2,+1}^{2}-H_{+3/2,+1}^{2}\Big{)}\\
&-2\hat{m_{\ell}}^{2}\Big{(}
\text{Re}[H_{+1/2,0}^{\dagger}H_{+1/2,t}]+\text{Re}[H_{-1/2,0}^{\dagger}H_{-1/2,t}]\\
&+\text{Re}[H_{+1/2,0}H_{+1/2,t}^{\dagger}]+\text{Re}[H_{-1/2,0}H_{-1/2,t}^{\dagger}]
\Big{)},
\end{split}
\end{equation}
\begin{equation}
\begin{split}
L_{3}=&-(1-\hat{m_{\ell}}^{2})\Big{(}H_{+1/2,0}^{2}+H_{-1/2,0}^{2}\Big{)}\\
&+\frac{1-\hat{m_{\ell}}^{2}}{2}\Big{(}H_{-3/2,-1}^{2}+H_{-1/2,-1}^{2}+H_{+1/2,+1}^{2}+H_{+3/2,+1}^{2}\Big{)}.
\end{split}
\end{equation}

The helicity amplitude $H_{\lambda^{\prime},\lambda_{W}}$ is defined as
\begin{equation}
H_{\lambda_{\mathcal{B}_{c}},\lambda_{W}}=\epsilon^{*}_{\mu}(\lambda_{W})
\langle\mathcal{B}_{c}(P^{\prime},\lambda_{\mathcal{B}_{c}})\vert V^{\mu}-A^{\mu}\vert\mathcal{B}_{b}(P,\lambda_{\mathcal{B}_{b}})\rangle
\end{equation}
with $\lambda$, $\lambda^{\prime}$, and $\lambda_{W}$ denoting the helicities of the initial state $\mathcal{B}_{b}$, the final state $\mathcal{B}_{c}$, and the off-shell $W$ boson, respectively. We have the relation $\lambda=\lambda^{\prime}-\lambda_{W}$. Their concerned expressions are ~\cite{Gutsche:2017wag}
\begin{equation}
\begin{split}
H^{V}_{1/2,t}=&\sqrt{\frac{2}{3}\frac{Q_{-}}{q^2}}\frac{Q_{+}}{2MM^{\prime}}
\bigg{(}g_1^V(q^2)M+g_2^V(q^2)M_{-}\\
&+g_3^V(q^2)\frac{M_{+}M_{-}-q^2}{2M^{\prime}}
+g_4^V(q^2)\frac{M_{+}M_{-}+q^2}{2M}\bigg{)},\\
H^{V}_{1/2,0}=&\sqrt{\frac{2}{3}\frac{Q_{+}}{q^2}}\bigg{(}
g_1^V(q^2)\frac{M_{+}M_{-}-q^2}{2M^{\prime}}+g_2^V(q^2)\frac{Q_{-}M_{+}}{2MM^{\prime}}\\
&+g_3^V(q^2)\frac{Q_{+}Q_{-}}{4MM^{\prime2}}
+g_4^V(q^2)\frac{Q_{+}Q_{-}}{4M^{2}M^{\prime}}\bigg{)},\\
H^{V}_{1/2,1}=&\sqrt{\frac{Q_{+}}{3}}\bigg{(}g_1^V(q^2)-g_2^V(q^2)\frac{Q_{-}}{MM^{\prime}}\bigg{)},\\
H^{V}_{3/2,1}=&\sqrt{Q_{+}}g_1^V(q^2),
\end{split}
\end{equation}
\begin{equation}
\begin{split}
H^{A}_{1/2,t}=&-\sqrt{\frac{2}{3}\frac{Q_{+}}{q^2}}\frac{Q_{-}}{2MM^{\prime}}
\bigg{(}f_1^A(q^2)M-f_2^A(q^2)M_{+}\\
&+f_3^A(q^2)\frac{M_{+}M_{-}-q^2}{2M^{\prime}}
+f_4^A(q^2)\frac{M_{+}M_{-}+q^2}{2M}\bigg{)},\\
H^{A}_{1/2,0}=&-\sqrt{\frac{2}{3}\frac{Q_{-}}{q^2}}\bigg{(}
f_1^A(q^2)\frac{M_{+}M_{-}-q^2}{2M^{\prime}}-f_2^A(q^2)\frac{Q_{+}M_{-}}{2MM^{\prime}}\\
&+f_3^A(q^2)\frac{Q_{+}Q_{-}}{4MM^{\prime2}}
+f_4^A(q^2)\frac{Q_{+}Q_{-}}{4M^2M^{\prime}}\bigg{)},\\
H^{A}_{1/2,1}=&\sqrt{\frac{Q_{-}}{3}}\bigg{(}f_1^A(q^2)-f_2^A(q^2)\frac{Q_{+}}{MM^{\prime}}\bigg{)},\\
H^{A}_{3/2,1}=&-\sqrt{Q_{-}}f_1^A(q^2),
\end{split}
\end{equation}
for the vector current and the axial-vector current, respectively, with $M_{\pm}=M\pm M^{\prime}$. The negative terms can be obtained by the relations
\begin{equation}
H^{V}_{-\lambda^{\prime},-\lambda_W}=+H^{V}_{\lambda^{\prime},\lambda_W},~~
H^{A}_{-\lambda^{\prime},-\lambda_W}=-H^{A}_{\lambda^{\prime},\lambda_W},
\end{equation}
and the total helicity amplitudes can be obtained by
\begin{equation}
H_{\lambda^{\prime},\lambda_W}=H^{V}_{\lambda^{\prime},\lambda_W}-H^{A}_{\lambda^{\prime},\lambda_W}.
\end{equation}

After performing the integral of the angle $\theta_{\ell}$, the differential decay width can be obtained by
\begin{equation}
\begin{split}
\frac{d\Gamma}{dq^2}=&\Big{\vert}\frac{G_{F}}{\sqrt{2}}V_{cb}\Big{\vert}^{2}
\frac{\sqrt{s_{+}s_{-}}q^{2}(1-\hat{m_{\ell}}^{2})^{2}}{512\pi^{3}M^{3}}
\Big{(}2L_{1}-\frac{2}{3}L_{3}\Big{)}.
\end{split}
\end{equation}
And then, the decay width can be obtained by carrying out the integral of $q^{2}$ in the range $m_{\ell}^{2}$ to $q^{2}_{\text{max}}$.

\begin{figure*}[htbp]\centering
  \begin{tabular}{lr}
  \includegraphics[width=80mm]{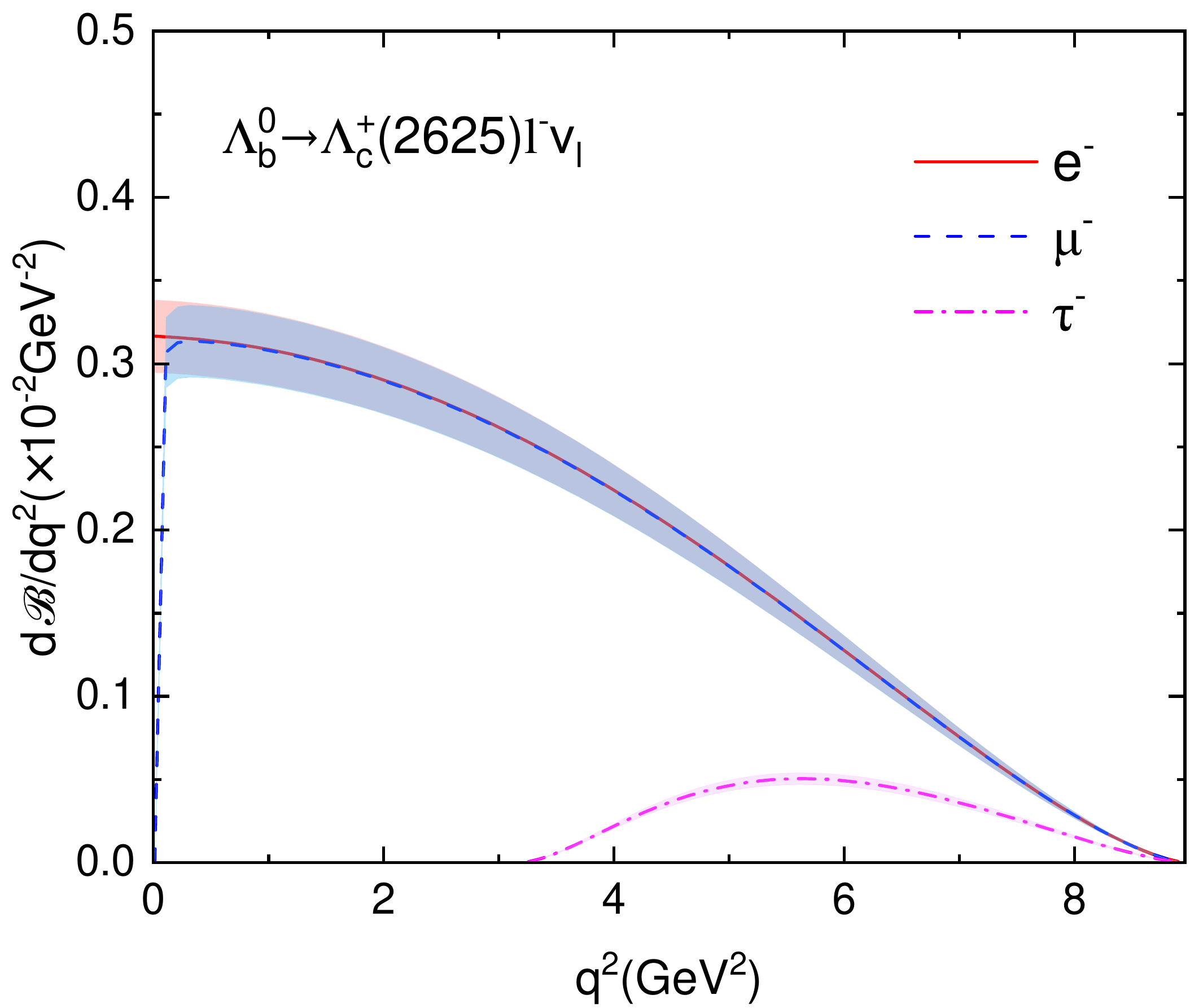}
  \includegraphics[width=80mm]{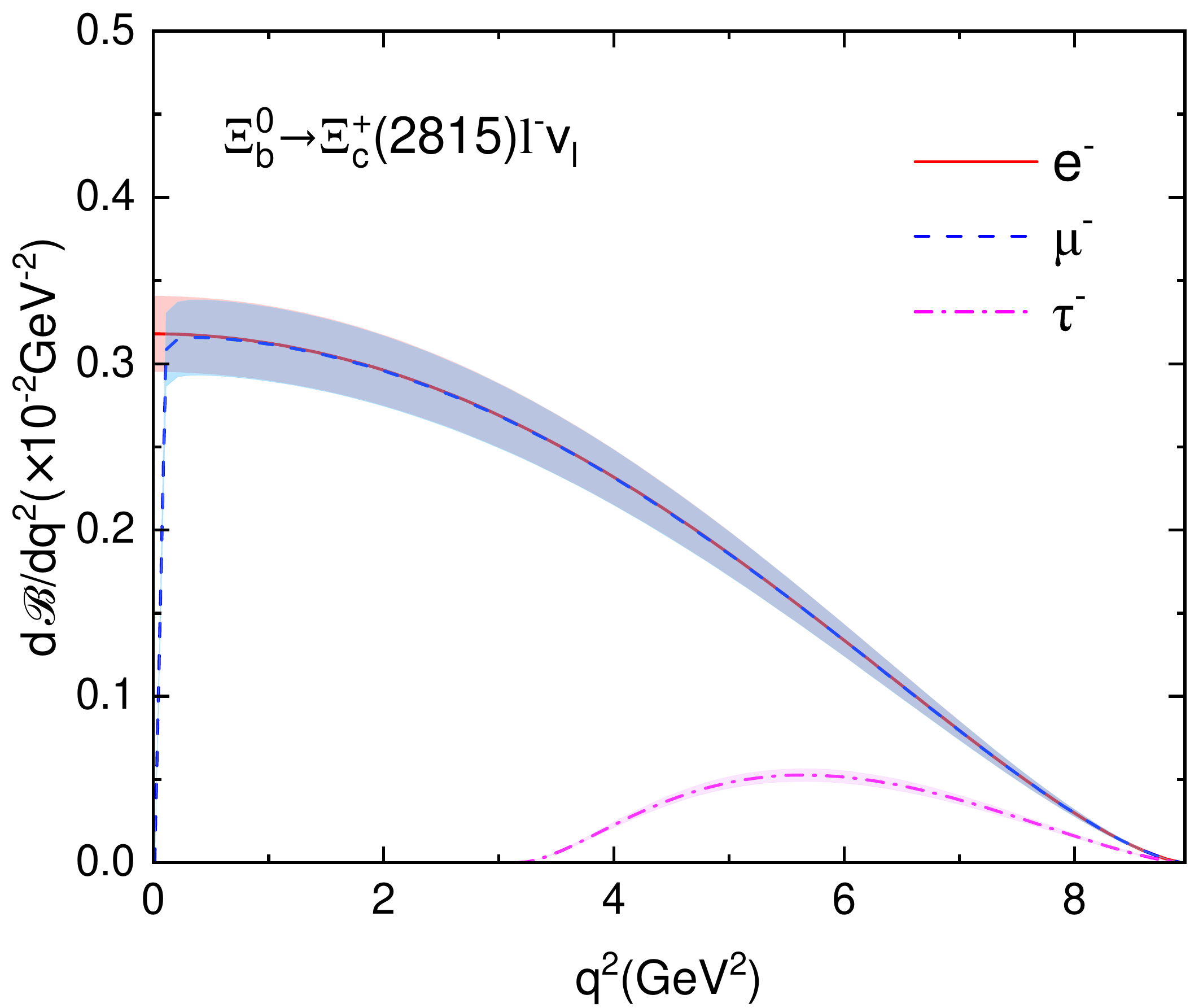}
  \end{tabular}
  \caption{The differential branching ratios of $\Lambda_b^0\to\Lambda_c^+(2625)\ell^-\nu_{\ell}$ and $\Xi_b^0\to\Xi_c^+(2815)\ell^-\nu_{\ell}$ with $\ell^{-}=e^{-},\mu^{-}, \text{or}\ \tau^{-}$.}
\label{fig:semilepton}
\end{figure*}

\begin{table*}[htbp]\centering
\caption{The comparison of our numerical results and the experimental measurement, as well as other theoretical results of the absolute branching ratios of $\Lambda_b^0\to\Lambda_c^+(2625)\ell^-\nu_{\ell}$ and $\Xi_b^{0,-}\to\Xi_c^{+,0}(2815)\ell^-\nu_{\ell}$ with $\ell=e,\mu,\tau$, where the branching ratios out of or in brackets in the second column correspond to the $\Xi_b^0\to\Xi_c(2815)^+$ and $\Xi_b^-\to\Xi_c(2815)^0$ transitions, respectively. Here, all values are given as a percent $(\%)$.}
\label{tab:semilepton}
\renewcommand\arraystretch{1.2}
\begin{tabular*}{172mm}{c@{\extracolsep{\fill}}ccccc}
\toprule[1pt]
\toprule[0.5pt]
$\text{Mode}$  &This work  &Expt.~\cite{ParticleDataGroup:2022pth}  &CCQM~\cite{Gutsche:2018nks}   &HQSS~\cite{Nieves:2019kdh}   &CQM~\cite{Pervin:2005ve}\\
\toprule[0.5pt]
$\Lambda_b^0\to\Lambda_c^+(2625)e^-\nu_{e}$   &$1.653\pm0.114$  &-  &$0.17\pm0.03$  &-  &$(0.88-1.40)$\\
$\Lambda_b^0\to\Lambda_c^+(2625)\mu^-\nu_{\mu}$   &$1.641\pm0.113$  &$1.3^{+0.6}_{-0.5}$  &$0.17\pm0.03$ &$3.5^{+1.3}_{-1.2}$  &$(0.88-1.40)$\\
$\Lambda_b^0\to\Lambda_c^+(2625)\tau^-\nu_{\tau}$   &$0.1688\pm0.0116$  &-  &$0.018\pm0.004$ &$0.38^{+0.09}_{-0.08}$  &$(0.18-0.22)$\\
\toprule[0.5pt]
$\Xi_b^{0(-)}\to\Xi_c^{+(0)}(2815)e^-\nu_{e}$   &$1.698\pm0.122\ (1.803\pm0.132)$  &-  &- & -  & -\\
$\Xi_b^{0(-)}\to\Xi_c^{+(0)}(2815)\mu^-\nu_{\mu}$   &$1.685\pm0.121\ (1.789\pm0.131)$  &-  &- & -  & -\\
$\Xi_b^{0(-)}\to\Xi_c^{+(0)}(2815)\tau^-\nu_{\tau}$   &$0.1758\pm0.0126\ (0.1868\pm0.0137)$  &-  &- & -  & -\\
\bottomrule[0.5pt]
\bottomrule[1pt]
\end{tabular*}
\end{table*}

Taking the form factors obtained by the light-front quark model as input, we calculate the semileptonic decays of the $\Lambda_b\to\Lambda_c(2625)$ and $\Xi_b\to\Xi_c(2815)$ processes. The masses of baryons and leptons are taken from the PDG~\cite{ParticleDataGroup:2022pth}, and the lifetimes of $\Lambda_b^0$ and $\Xi_{b}^{-,0}$ are fixed to be
\begin{equation*}
\begin{split}
\tau_{\Lambda_b^0}=&(1.471\pm0.009)\ \text{fs},\\
\tau_{\Xi_b^-}=&(1.572\pm0.040)\ \text{fs},\\
\tau_{\Xi_b^0}=&(1.480\pm0.030)\ \text{fs},
\end{split}
\end{equation*}
respectively, averaged by the PDG~\cite{ParticleDataGroup:2022pth}. Besides, the involved CKM matrix element is $V_{cb}=(40.8\pm1.4)\times10^{-3}$~\cite{ParticleDataGroup:2022pth}.

The $q^2$ dependence of the differential branching ratios are shown in Fig.~\ref{fig:semilepton}. Since the ones of $\Xi_b^-\to\Xi_c^0(2815)\ell^-\nu_{\ell}$ act similar with the neutral one, we would not display them here.
In the meantime, we also present the branching ratios, and compare our results with the experimental data and other theoretical results, including the CCQM~\cite{Gutsche:2018nks}, the HQSS~\cite{Nieves:2019kdh}, and the constituent quark model (CQM)~\cite{Pervin:2005ve}, in Table~\ref{tab:semilepton}.

Obviously, our result $\mathcal{B}(\Lambda_b^0\to\Lambda_c^+(2625)\mu^-\nu_{\mu})=(1.617\pm0.003)\%$ is consistent with the current experimental data $(1.3^{+0.6}_{-0.5})\%$~\cite{ParticleDataGroup:2022pth}. Moreover, the predicted branching ratios of the electron and muon channels can reach up to the magnitude of $1\%$, which are accessible at LHCb. From Table~\ref{tab:semilepton}, we notice that our results of $\Lambda_b\to\Lambda_c(2625)\ell^{-}\nu_{\ell}$ are consistent with the estimate from the HQSS~\cite{Nieves:2019kdh} and the CQM~\cite{Pervin:2005ve}, but are larger than the CCQM results~\cite{Gutsche:2018nks}. Besides, we also find that there exists difference of our result of $\mathcal{B}(\Xi_b\to\Xi_c(2815)\ell^{-}\nu_{\ell})$ and that given by Ref.~\cite{Pavao:2017cpt}, where the $\Xi_c(2815)$ resonance is assumed as the dynamically effect~\cite{Pavao:2017cpt}. It shows that the branching ratios of these discussed transitions are dependent on the different structure assignments to the $\Lambda_c(2625)$ and
$\Xi_c(2815)$.
We expect more theoretical studies and the ongoing experiment to explore them, which will be a crucial test to our result. What is more important is that different structure assignments to the $\Lambda_c(2625)$ and
$\Xi_c(2815)$ can be further distinguished.

Additionally, other important physical observables, including the leptonic forward-backward asymmetry ($A_{FB}$), the final hadron polarization ($P_{B}$), and the lepton polarization ($P_{\ell}$) are also investigated in this work. The leptonic forward-backward asymmetry $A_{FB}$ can be obtained by
\begin{equation}
\begin{split}
A_{FB}(q^2)=\frac{\Big{(}\int_{0}^{1}-\int_{-1}^{0}\Big{)}d\cos\theta_{\ell}\frac{d^{2}\Gamma}{dq^{2}d\cos\theta_{\ell}}}
{\Big{(}\int_{0}^{1}+\int_{-1}^{0}\Big{)}d\cos\theta_{\ell}\frac{d^{2}\Gamma}{dq^{2}d\cos\theta_{\ell}}}=
\frac{3L_{2}}{6L_{1}-2L_{3}}.
\end{split}
\end{equation}
The final hadron polarization $P_{B}$ is defined as
\begin{equation}
\begin{split}
P_{B}(q^2)=\frac{d\Gamma^{\lambda^{\prime}=(+3/2,+1/2)}/dq^2-d\Gamma^{\lambda^{\prime}=(-3/2,-1/2)}/dq^2}{d\Gamma/dq^2}
\end{split}
\end{equation}
with $\lambda^{\prime}$ representing the polarization of the final charmed hadron, and
\begin{equation}
\begin{split}
\frac{d\Gamma^{\lambda^{\prime}=(+3/2,+1/2)}}{dq^2}=&
\frac{4}{3}
\Bigg{(}(2+\hat{m_{\ell}}^{2})\Big{(}H_{1/2,0}^{2}+H_{1/2,1}^{2}+H_{3/2,1}^{2}\Big{)}\\
&+3\hat{m_{\ell}}^{2}H_{1/2,t}^{2}\Bigg{)},
\end{split}
\end{equation}
\begin{equation}
\begin{split}
\frac{d\Gamma^{\lambda^{\prime}=(-3/2,-1/2)}}{dq^2}=&
\frac{4}{3}
\Bigg{(}(2+\hat{m_{\ell}}^{2})\Big{(}H_{-1/2,0}^{2}+H_{-1/2,-1}^{2}\\
&+H_{-3/2,-1}^{2}\Big{)}+3\hat{m_{\ell}}^{2}H_{-1/2,t}^{2}\Bigg{)}.
\end{split}
\end{equation}
And the lepton polarization $P_{\ell}$ can be obtained by
\begin{equation}
\begin{split}
P_{\ell}(q^2)=\frac{d\Gamma^{\lambda_{\ell}=+1/2}/dq^2-d\Gamma^{\lambda_{\ell}=-1/2}/dq^2}{d\Gamma/dq^2},
\end{split}
\end{equation}
with $\lambda_{\ell}$ denoting the polarization of the lepton $\ell^{-}$, and
\begin{equation}
\begin{split}
\frac{d\Gamma^{\lambda_{\ell}=+1/2}}{dq^2}=&
\frac{4}{3}\hat{m_{\ell}}^{2}
\Bigg{(}H_{+1/2,0}^{2}+H_{-1/2,0}^{2}+H_{1/2,1}^{2}+H_{3/2,1}^{2}\\
&+H_{-1/2,-1}^{2}+H_{-3/2,-1}^{2}
+3H_{1/2,t}^{2}+3H_{-1/2,t}^{2}
\Bigg{)},
\end{split}
\end{equation}
\begin{equation}
\begin{split}
\frac{d\Gamma^{\lambda_{\ell}=-1/2}}{dq^2}=&
\frac{8}{3}
\Bigg{(}H_{1/2,0}^{2}+H_{-1/2,0}^{2}+H_{1/2,1}^{2}+H_{3/2,1}^{2}\\
&+H_{-1/2,-1}^{2}+H_{-3/2,-1}^{2}
\Bigg{)}.
\end{split}
\end{equation}
Here, we neglect the common term
\begin{equation}
\begin{split}
\Big{\vert}\frac{G_{F}}{\sqrt{2}}V_{cb}\Big{\vert}^{2}
\frac{\sqrt{s_{+}s_{-}}q^{2}(1-\hat{m_{\ell}}^{2})^{2}}{512\pi^{3}M^{3}}
\end{split}
\end{equation}
for abbreviation.

\begin{figure*}[htbp]\centering
  \begin{tabular}{lr}
  \includegraphics[width=80mm]{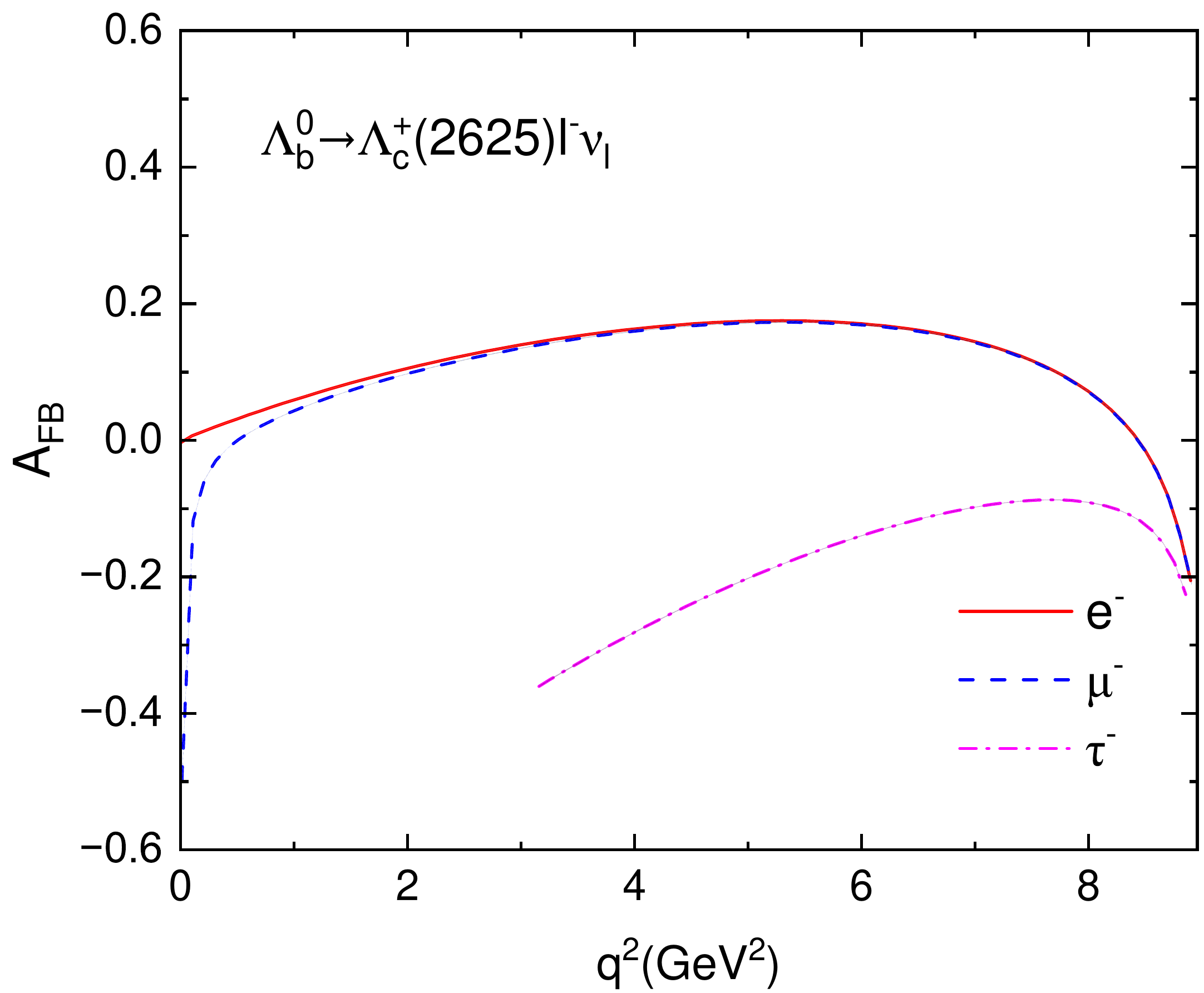}
  \includegraphics[width=80mm]{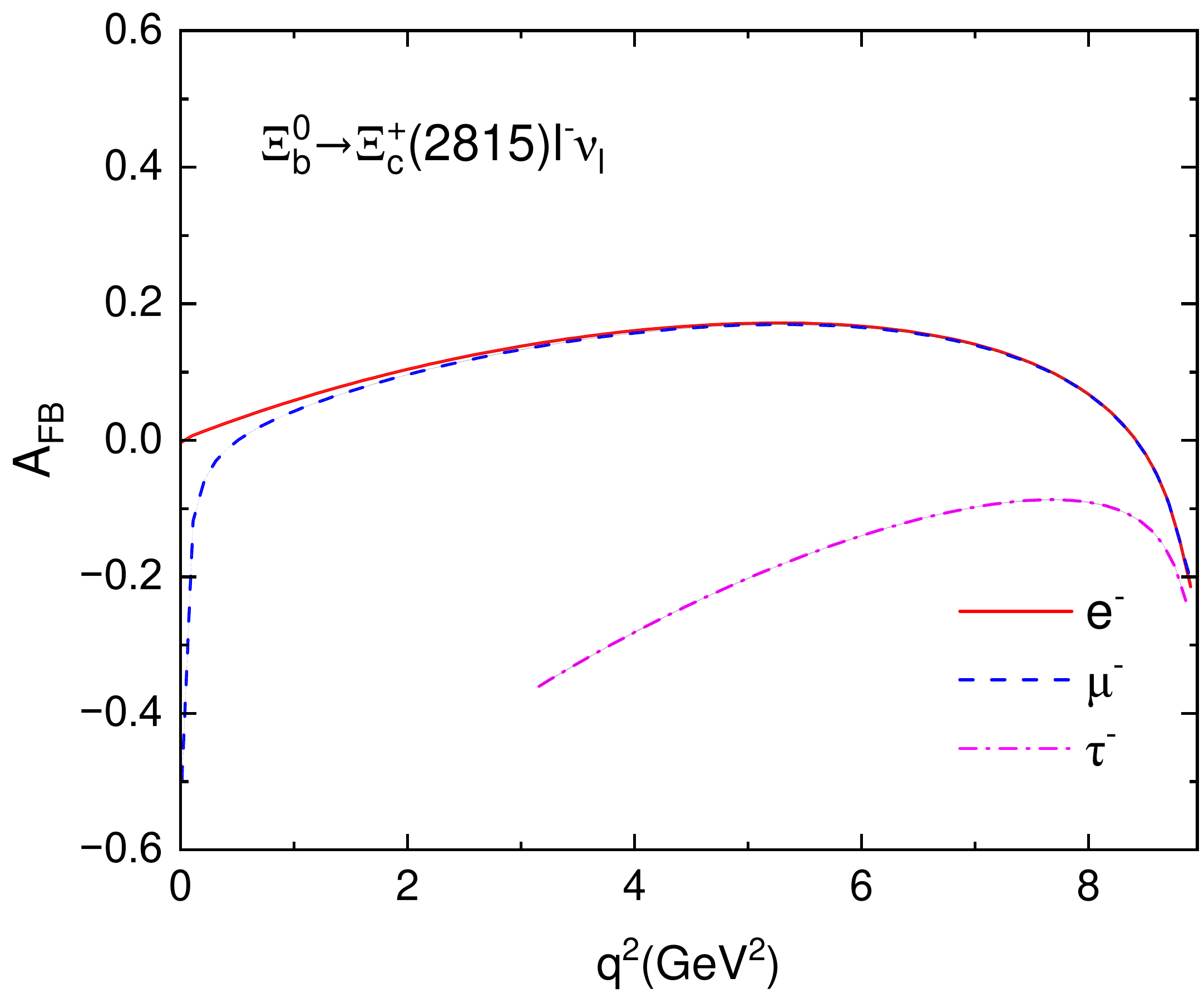}
  \end{tabular}
  \caption{The leptonic forward-backward asymmetries ($A_{FB}$) of $\Lambda_b^0\to\Lambda_c^+(2625)\ell^-\nu_{\ell}$ and $\Xi_b^0\to\Xi_c^+(2815)\ell^-\nu_{\ell}$ with $\ell^{-}=e^{-},\mu^{-}, \text{or}\ \tau^{-}$. Here, the uncertainties are also added. However, they are not obvious when we present the corresponding results.}
\label{fig:AFB}
\end{figure*}

\begin{figure*}[htbp]\centering
  \begin{tabular}{lr}
  \includegraphics[width=80mm]{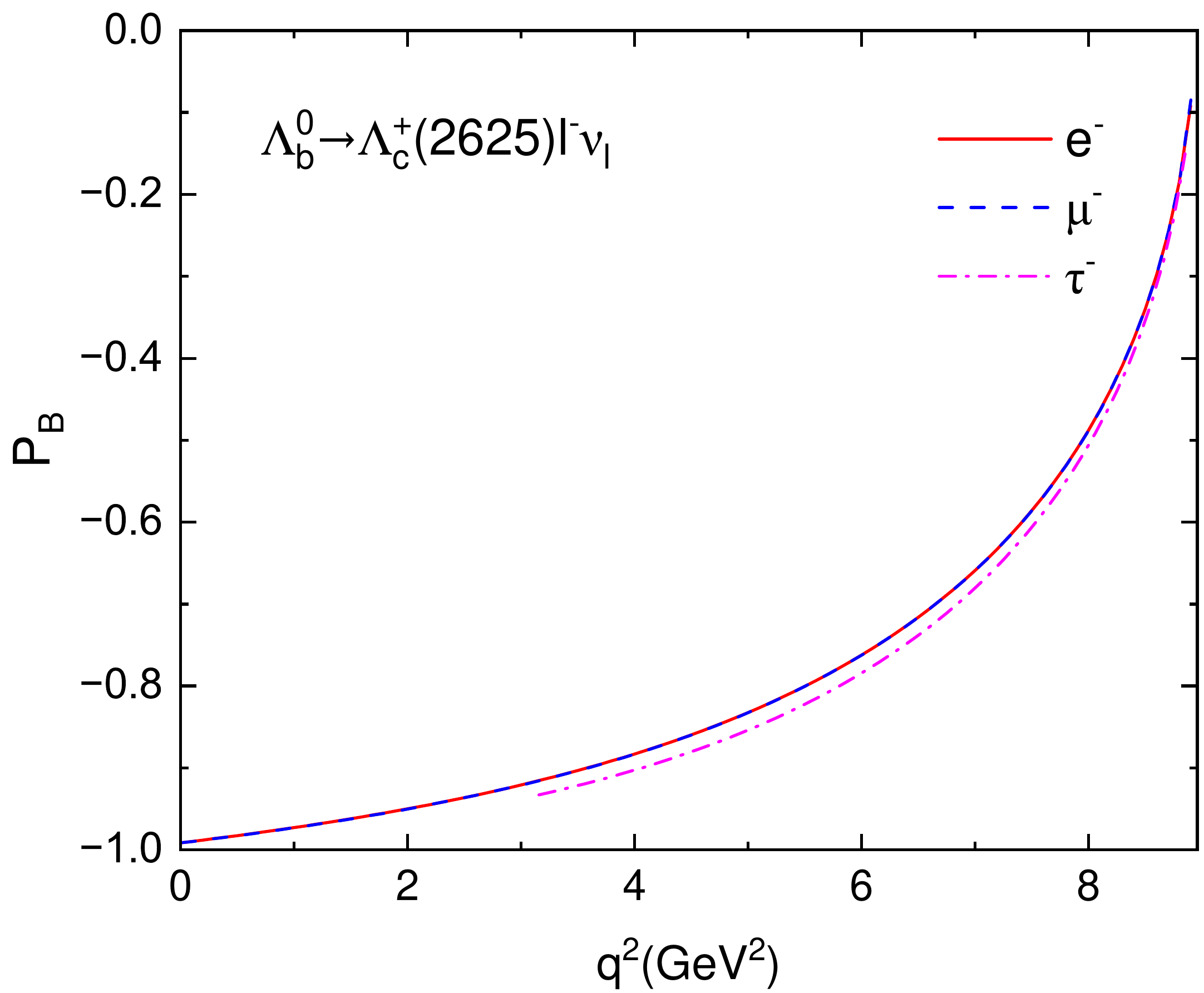}
  \includegraphics[width=80mm]{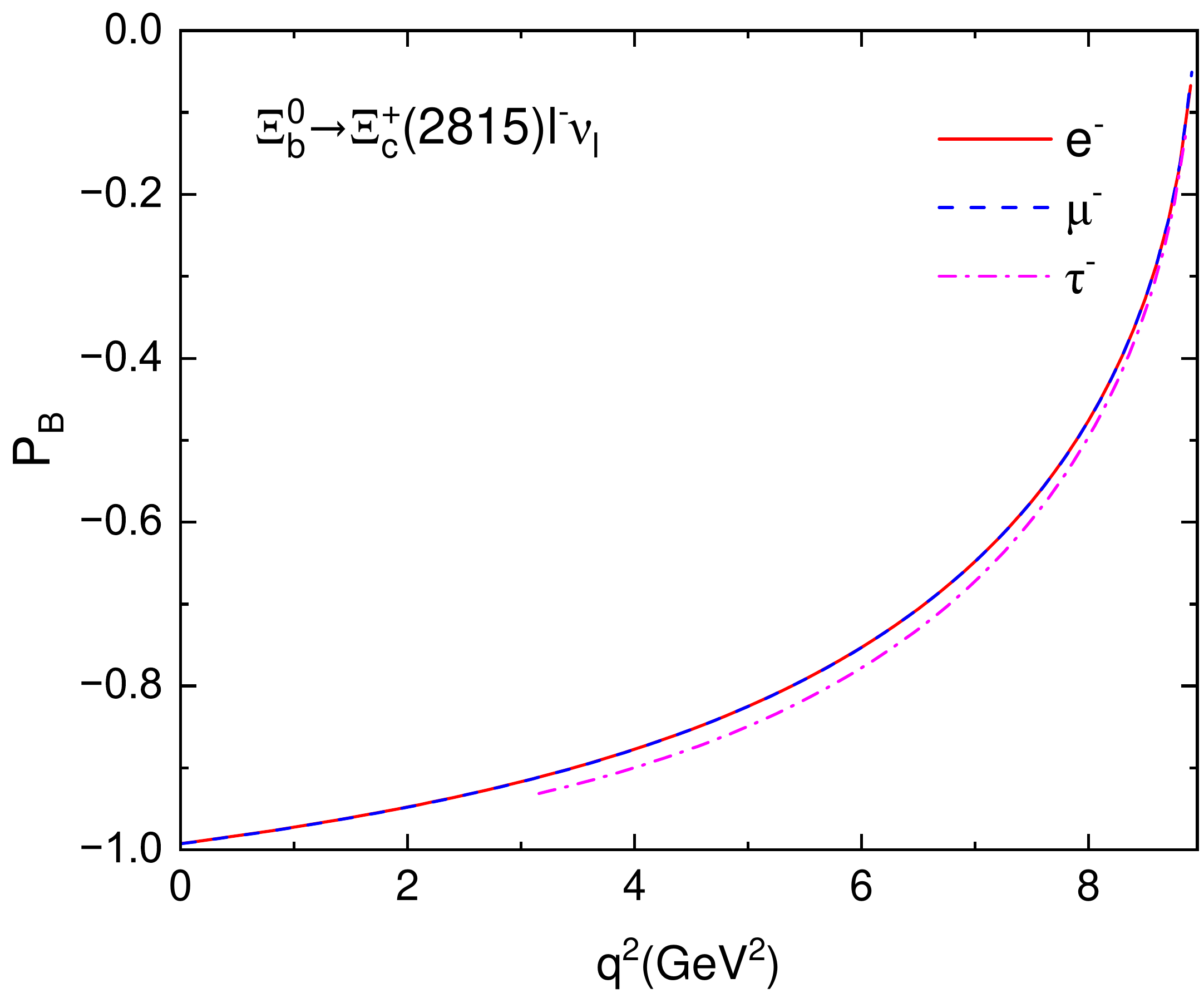}
  \end{tabular}
  \caption{The final hadron polarizations ($P_{B}$) of $\Lambda_b^0\to\Lambda_c^+(2625)\ell^-\nu_{\ell}$ and $\Xi_b^0\to\Xi_c^+(2815)\ell^-\nu_{\ell}$ with $\ell^{-}=e^{-},\mu^{-}, \text{or}\ \tau^{-}$. Here, the uncertainties are also added. However, they are not obvious when we present the corresponding results.}
\label{fig:PB}
\end{figure*}

\begin{figure*}[htbp]\centering
  \begin{tabular}{lr}
  \includegraphics[width=80mm]{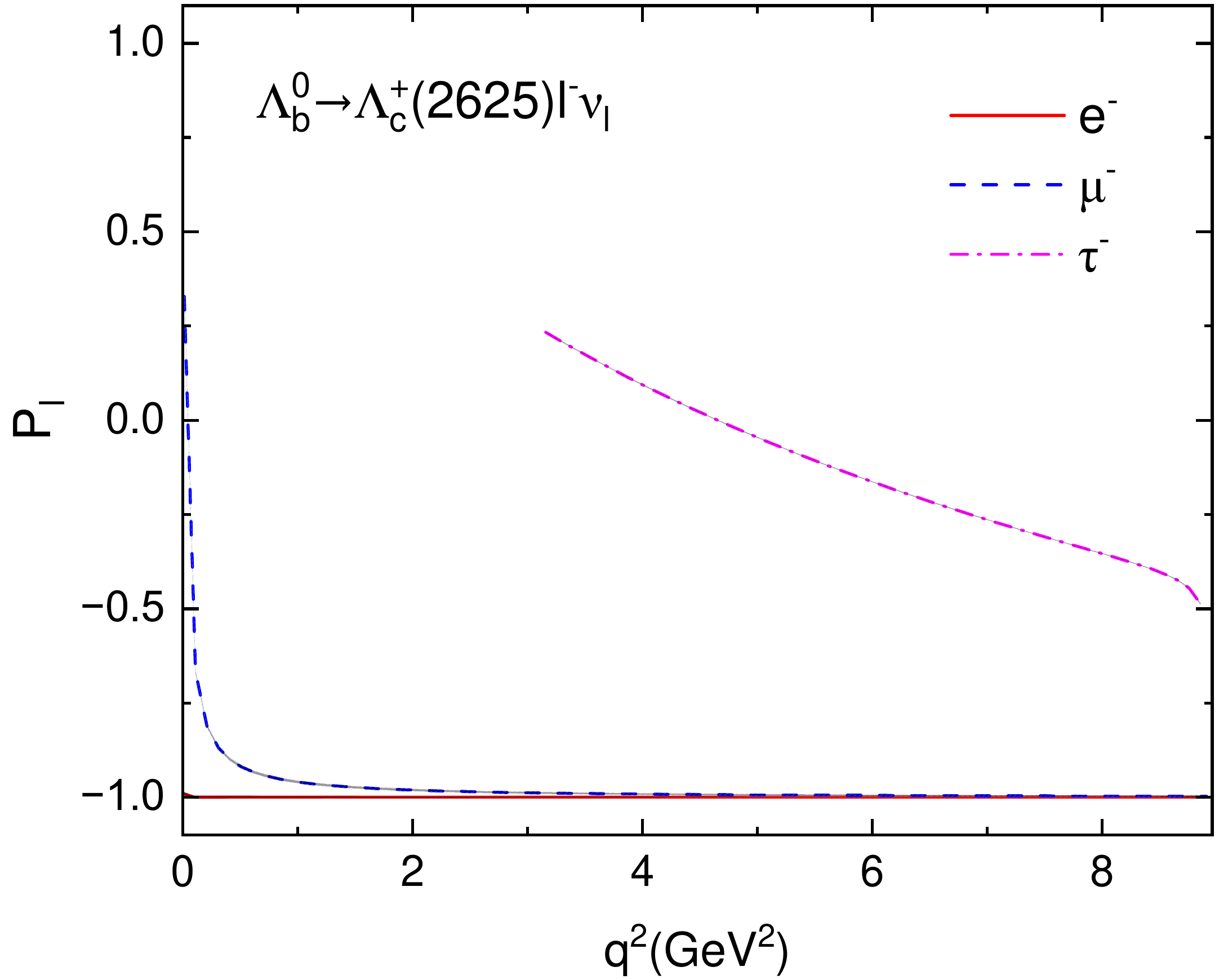}
  \includegraphics[width=80mm]{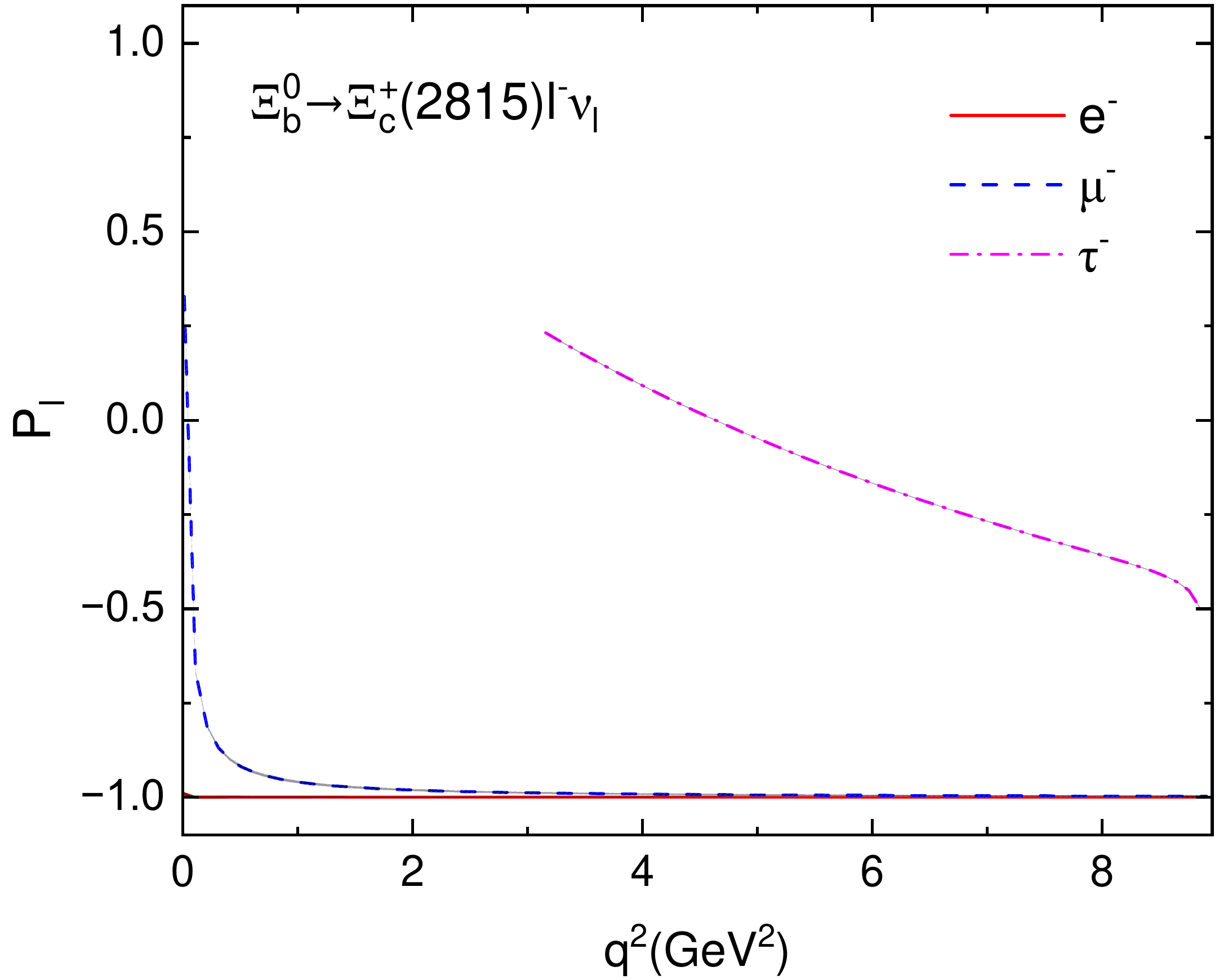}
  \end{tabular}
  \caption{The lepton polarizations ($P_{\ell}$) of $\Lambda_b^0\to\Lambda_c^+(2625)\ell^-\nu_{\ell}$ and $\Xi_b^0\to\Xi_c^+(2815)\ell^-\nu_{\ell}$ with $\ell^{-}=e^{-},\mu^{-}, \text{or}\ \tau^{-}$. Here, the uncertainties are also added. However, they are not obvious when we present the corresponding results.}
\label{fig:Pl}
\end{figure*}

The $q^2$ dependence of the leptonic forward-backward asymmetries ($A_{FB}$), the final hadron polarizations ($P_{B}$), and the lepton polarizations ($P_{\ell}$) of the concerned semileptonic decays is displayed in Fig.~\ref{fig:AFB}, Fig.~\ref{fig:PB}, and Fig.~\ref{fig:Pl}, respectively. The future experimental measurement of these physical observables may provide valuable information to these discussed weak decays.


Moreover, we are also interested in the ratios of branching fractions
\begin{eqnarray*}
R_{\Lambda_c(2625)}=\frac{\mathcal{B}(\Lambda_b\to\Lambda_c(2625)\tau^{-}\nu_{\tau})}{\mathcal{B}(\Lambda_b\to\Lambda_c(2625)\ell^{-}\nu_{\ell})}\approx 0.10,\\
R_{\Xi_c(2815)}=\frac{\mathcal{B}(\Xi_b\to\Xi_c(2815)\tau^{-}\nu_{\tau})}{\mathcal{B}(\Xi_b\to\Xi_c(2815)\ell^{-}\nu_{\ell})}\approx 0.10
\end{eqnarray*}
with $\ell^{-}=e^{-}\ \text{or}\ \mu^{-}$, which reflect the LFU. Our result of $R_{\Lambda_c(2625)}$ is also  consistent with $0.11\pm0.02$ estimated by the CCQM~\cite{Gutsche:2018nks}.

\subsection{The color-allowed two-body nonleptonic decays}

In this subsection, we further evaluate the color-allowed two-body nonleptonic decays $\Lambda_b^0\to\Lambda_c^+(2625)M^-$ and $\Xi_b^{0,-}\to\Xi_c^{+,0}(2815)M^-$ with $M^-$ being a pseudoscalar meson ($\pi^-$, $K^-$, $D^-$, or $D_s^-$) or a vector meson ($\rho^-$, $K^{*-}$, $D^{*-}$, or $D_s^{*-}$). Based on the na\"{i}ve factorization assumption, the hadronic transition matrix element can be factorized into a product of two independent matrix elements
\begin{equation}
\begin{split}
\langle\mathcal{B}_c(P^{\prime},&J_{z}^{\prime})M^{-}\vert\mathcal{H}_{\text{eff}}\vert\mathcal{B}_b(P,J_{z})\rangle\\
=&\frac{G_F}{\sqrt{2}} V_{cb} V_{qq^{\prime}}^{\ast}
\langle M^{-}|\bar{q}^{\prime}\gamma_{\mu}(1-\gamma_{5})q|0\rangle\\
&\times
\langle\mathcal{B}_c(P^{\prime},J_{z}^{\prime})|\bar{c}\gamma^{\mu}(1-\gamma_{5})b|\mathcal{B}_b(P,J_{z})\rangle,
\end{split}
\end{equation}
where the meson part is determined by a decay parameter as
\begin{equation}
\begin{split}
\langle M|\bar{q}^{\prime}&\gamma_{\mu}(1-\gamma_{5})q|0\rangle
=\left\{
\begin{array}{ll}
if_{P}q_{\mu}, &M \in \text{pseudoscalar\ meson}\\
f_{V}\epsilon_{\mu}^{*}m_{V}, &M \in \text{vector\ meson}
\end{array}.
\right.
\end{split}
\end{equation}

{Frankly speaking, the na\"{i}ve factorization assumption works well for the color-allowed dominated decays. However, there exists the case, where the color-suppressed and penguin dominated processes can not be explained by the na\"{i}ve factorization, which may show important nonfactorizable contribution to nonleptonic decays \cite{Zhu:2018jet}. As shown in Refs.~\cite{Lu:2009cm,Chua:2018lfa,Chua:2019yqh}, the nonfactorizable contributions in bottom baryon decays are cosiderable comparing with the factorized ones. But the precise study of nonfactorizable contributions is beyond the scope of the present work, we still take the approximation of using the na\"{i}ve factorization assumption.}

In our calculation, the decay constants of these involved pseudoscalar and vector mesons include~\cite{Cheng:2003sm,Chua:2019yqh,Li:2021kfb}
\begin{equation*}
\begin{split}
&f_{\pi}=130.2,\ f_{K}=155.6,\ f_{D}=211.9,\ f_{D_s}=249.0,\\
&f_{\rho}=216,\ f_{K^*}=210,\ f_{D^*}=220,\ f_{D_s^*}=230,
\end{split}
\end{equation*}
in the unit of MeV.

On the other hand, the decay amplitudes of the $\mathcal{B}_{b}\to\mathcal{B}_{c}P$ and $\mathcal{B}_{b}\to\mathcal{B}_{c}V$ processes can be parameterized as
\begin{equation}
\mathcal{A}\big{[}\mathcal{B}_{b}\to\mathcal{B}_{c}P\big{]}=iq_{\mu}\bar{u}^{\mu}(C+D\gamma_5)u,
\end{equation}
\begin{equation}
\begin{split}
\mathcal{A}\big{[}\mathcal{B}_{b}\to\mathcal{B}_{c}V\big{]}=&\epsilon^{\ast\mu}
\bar{u}^{\nu}\big{[}
g_{\nu\mu}(C_1+D_1\gamma_5)+q_{\nu}\gamma_{\mu}(C_2+D_2\gamma_5)\\
&+q_{\nu}P_{\mu}(C_3+D_3\gamma_5)
\big{]}u,
\end{split}
\end{equation}
respectively, with $P(V)$ denoting the pseudoscalar (vector) meson, where the parity-violated and parity-conserved amplitudes are written as
\begin{equation}
\begin{split}
C=&\frac{G_F}{\sqrt{2}}V_{cb}V_{qq^{\prime}}a_{1}f_{P}\bigg{[}
g_{1}^{V}(m_{P}^2)+(M-M^{\prime})\frac{g_{2}^{V}(m_{P}^2)}{M}\\
&+\frac{1}{2}(M^2-M^{\prime2}-m_{P}^2)\big{(}\frac{g_{3}^{V}(m_{P}^2)}{MM^{\prime}}+\frac{g_{4}^{V}(m_{P}^2)}{M^2}\big{)}
-m_{P}^2\frac{g_{3}^{V}(m_{P}^2)}{MM^{\prime}}
\bigg{]},\\
D=&-\frac{G_F}{\sqrt{2}}V_{cb}V_{qq^{\prime}}a_{1}f_{P}\bigg{[}
f_{1}^{A}(m_{P}^2)-(M+M^{\prime})\frac{f_{2}^{A}(m_{P}^2)}{M}\\
&+\frac{1}{2}(M^2-M^{\prime2}-m_{P}^2)\big{(}\frac{f_{3}^{A}(m_{P}^2)}{MM^{\prime}}+\frac{f_{4}^{A}(m_{P}^2)}{M^2}\big{)}
-m_{P}^2\frac{g_{3}^{V}(m_{P}^2)}{MM^{\prime}}
\bigg{]},\\
\end{split}
\end{equation}

\begin{equation}
\begin{split}
C_1=&\frac{G_F}{\sqrt{2}}V_{cb}V_{qq^{\prime}}a_{1}f_{V}g_{1}^{V}(m_{V}^2),\\
D_1=&-\frac{G_F}{\sqrt{2}}V_{cb}V_{qq^{\prime}}a_{1}f_{V}f_{1}^{A}(m_{V}^2),\\
C_2=&\frac{G_F}{\sqrt{2}}V_{cb}V_{qq^{\prime}}a_{1}f_{V}\frac{g_{2}^{V}(m_{V}^2)}{M},\\
D_2=&-\frac{G_F}{\sqrt{2}}V_{cb}V_{qq^{\prime}}a_{1}f_{V}\frac{f_{2}^{A}(m_{V}^2)}{M},\\
C_3=&\frac{G_F}{\sqrt{2}}V_{cb}V_{qq^{\prime}}a_{1}f_{V}\bigg{(}\frac{g_{3}^{V}(m_{V}^2)}{MM^{\prime}}+\frac{g_{4}^{V}(m_{V}^2)}{M^2}\bigg{)},\\
D_3=&-\frac{G_F}{\sqrt{2}}V_{cb}V_{qq^{\prime}}a_{1}f_{V}\bigg{(}\frac{f_{3}^{A}(m_{V}^2)}{MM^{\prime}}+\frac{f_{4}^{A}(m_{V}^2)}{M^2}\bigg{)}.
\end{split}
\end{equation}
The $m_{P}(m_{V})$ is the mass of the emitted pseudoscalar (vector) meson, and $a_{1}=c_{1}+c_{2}/N\approx1.018$~\cite{Chua:2019yqh}. Besides, the CKM matrix elements are~\cite{ParticleDataGroup:2022pth}
\begin{equation*}
\begin{split}
&V_{cb}=(40.8\pm1.4)\times10^{-3},\ V_{ud}=0.97373\pm0.00031,\\
&V_{us}=0.2243\pm0.0008,\ V_{cd}=0.221\pm0.004,\\
&V_{cs}=0.975\pm0.006.
\end{split}
\end{equation*}

Finally, the decay width and asymmetry parameter can be evaluated by
\begin{equation}
\begin{split}
\Gamma=&\frac{\vert\vec{p}_{c}\vert^3}{12\pi}\bigg{[}\frac{(M+M^{\prime})^2-m_{P}^2}{M^{\prime2}}\vert C\vert^2+\frac{(M-M^{\prime})^2-m_{P}^2}{M^{\prime2}}\vert D\vert^2\bigg{]},\\
\alpha=&-\frac{2\kappa \text{Re}[C^{\ast}D]}{\vert C\vert^2+\kappa^2\vert D\vert^2}
\label{eq:physicsP}
\end{split}
\end{equation}
with $\kappa=\vert\vec{p}_c\vert/(E^{\prime}+M^{\prime})$, and
\begin{equation}
\begin{split}
\Gamma=&\frac{\vert\vec{p}_c\vert}{32\pi M^2}\sum_{\lambda_V}
\Big{(}\vert h_{\lambda_V+1/2,\lambda_V;1/2}^{\text{PV}} \vert^2+\vert h_{\lambda_V+1/2,\lambda_V;1/2}^{\text{PC}} \vert^2\Big{)},\\
\alpha=&\frac{\sum_{\lambda_V}2h_{\lambda_V+1/2,\lambda_V;1/2}^{\text{PV}}h_{\lambda_V+1/2,\lambda_V;1/2}^{\text{PC}}}
{\sum_{\lambda_V}(\vert h_{\lambda_V+1/2,\lambda_V;1/2}^{\text{PV}} \vert^2+\vert h_{\lambda_V+1/2,\lambda_V;1/2}^{\text{PC}} \vert^2)}
\label{eq:physicsV}
\end{split}
\end{equation}
with
\begin{equation}
\begin{split}
h_{3/2,1;1/2}^{\text{PV(PC)}}=&\mp\sqrt{2s_{\pm}}C_{1}(D_{1}),\\
h_{-1/2,-1;1/2}^{\text{PV(PC)}}=&\mp\sqrt{\frac{2s_{\pm}}{3}}\Big{[}C_{1}(D_{1})-\frac{s_{\mp}}{M^{\prime}}C_{2}(D_{2})\Big{]},\\
h_{1/2,0;1/2}^{\text{PV(PC)}}=&\mp\frac{\sqrt{s_{\pm}}}{2\sqrt{3}M^{\prime}m_{V}}\Big{[}
2(M^2-M^{\prime2}-m_{V}^2)C_{1}(D_{1})\\
&\pm2s_{\mp}(M\pm M^{\prime})C_{2}(D_{2})+s_{+}s_{-}C_{3}(D_{3})
\Big{]},
\end{split}
\end{equation}
for the cases associated with the pseudoscalar and the vector meson emitted processes, respectively. The $\vec{p}_{c}$ is the three-momentum of the daughter baryon (or meson) in the rest frame of the parent baryon, while the $M(M^{\prime})$ is the mass of parent (daughter) baryon and $E^{\prime}$ denotes the energy of the daughter baryon.

\begin{table*}[htbp]\centering
\caption{The absolute branching ratios and up-down asymmetry parameters of the $\Lambda_b^0\to\Lambda_c^+(2625)M^-$ decays with $M$ denoting a pseudoscalar or vector meson. We also compare the branching ratios (in the unit of $10^{-3}$) with those given by Ref.~\cite{Chua:2019yqh} in the fourth column.}
\label{tab:nonleptonLambdab}
\renewcommand\arraystretch{1.05}
\begin{tabular*}{152mm}{l@{\extracolsep{\fill}}ccc}
\toprule[1pt]
\toprule[0.5pt]
Mode     &$\mathcal{B}\ (\times10^{-3})$    &$\alpha$   &Ref.~\cite{Chua:2019yqh}\\
\midrule[0.5pt]
$\Lambda_b^{0}\to\Lambda_c(2625)^{+}\pi^-$        &$3.12\pm0.15$  &$-0.99\pm0.07$
&$2.40^{+4.09}_{-1.82}$\\
$\Lambda_b^{0}\to\Lambda_c(2625)^{+}\rho^-$       &$4.25\pm0.23$  &$-0.88\pm0.07$
&$4.38^{+6.78}_{-3.17}$\\
$\Lambda_b^{0}\to\Lambda_c(2625)^{+}K^-$          &$0.232\pm0.012$  &$-0.99\pm0.07$
&$0.17^{+0.30}_{-0.13}$\\
$\Lambda_b^{0}\to\Lambda_c(2625)^{+}K^{*-}$       &$0.212\pm0.011$  &$-0.85\pm0.07$
&$0.22^{+0.33}_{-0.16}$\\
$\Lambda_b^{0}\to\Lambda_c(2625)^{+}D^-$          &$0.266\pm0.016$  &$-0.92\pm0.07$
&$0.13^{+0.22}_{-0.10}$\\
$\Lambda_b^{0}\to\Lambda_c(2625)^{+}D^{*-}$       &$0.161\pm0.007$  &$-0.45\pm0.05$
&$0.13^{+0.17}_{-0.08}$\\
$\Lambda_b^{0}\to\Lambda_c(2625)^{+}D_s^-$        &$6.60\pm0.40$  &$-0.90\pm0.07$
&$2.88^{+4.92}_{-2.16}$\\
$\Lambda_b^{0}\to\Lambda_c(2625)^{+}D_s^{*-}$     &$3.15\pm0.13$  &$-0.41\pm0.04$
&$2.41^{+2.98}_{-1.52}$\\
\bottomrule[0.5pt]
\bottomrule[1pt]
\end{tabular*}
\end{table*}

\begin{table*}[htbp]\centering
\caption{The absolute branching ratios and up-down asymmetry parameters of the $\Xi_b^{0,-}\to\Xi_c^{+,0}(2815)M^-$ decays with $M$ denoting a pseudoscalar or vector meson, where the branching ratios out of or in brackets correspond to the $\Xi_{b}^{0}\to\Xi_{c}(2815)^{+}$ and $\Xi_{b}^{-}\to\Xi_{c}(2815)^{0}$ transitions, respectively. We also compare the branching ratios (in the unit of $10^{-3}$) with those given by Ref.~\cite{Chua:2019yqh} in the fourth column.}
\label{tab:nonleptonXib}
\renewcommand\arraystretch{1.05}
\begin{tabular*}{152mm}{l@{\extracolsep{\fill}}ccc}
\toprule[1pt]
\toprule[0.5pt]
Mode     &$\mathcal{B}\ (\times10^{-3})$    &$\alpha$  &Ref.~\cite{Chua:2019yqh}\\
\midrule[0.5pt]
$\Xi_b^{0,-}\to\Xi_c(2815)^{+,0}\pi^-$           &$3.13\pm0.17\ (3.33\pm0.18)$    &$-0.99\pm0.07$
&$3.32^{+6.08}_{-2.63}\ (3.53^{+6.46}_{-2.80})$\\
$\Xi_b^{0,-}\to\Xi_c(2815)^{+,0}\rho^-$          &$4.29\pm0.25\ (4.55\pm0.27)$    &$-0.88\pm0.07$
&$6.10^{+9.95}_{-4.55}\ (6.49^{+10.58}_{-4.84})$\\
$\Xi_b^{0,-}\to\Xi_c(2815)^{+,0}K^-$             &$0.233\pm0.012\ (0.248\pm0.014)$    &$-0.99\pm0.07$
&$0.24^{+0.44}_{-0.19}\ (0.26^{+0.47}_{-0.20})$\\
$\Xi_b^{0,-}\to\Xi_c(2815)^{+,0}K^{\ast-}$       &$0.214\pm0.012\ (0.227\pm0.013)$    &$-0.85\pm0.07$
&$0.30^{+0.48}_{-0.22}\ (0.32^{+0.51}_{-0.24})$\\
$\Xi_b^{0,-}\to\Xi_c(2815)^{+,0}D^-$             &$0.275\pm0.017\ (0.292\pm0.019)$    &$-0.92\pm0.07$
&$0.19^{+0.33}_{-0.14}\ (0.20^{+0.35}_{-0.15})$\\
$\Xi_b^{0,-}\to\Xi_c(2815)^{+,0}D^{\ast-}$       &$0.167\pm0.008\ (0.177\pm0.009)$    &$-0.45\pm0.05$
&$0.19^{+0.24}_{-0.12}\ (0.20^{+0.26}_{-0.13})$\\
$\Xi_b^{0,-}\to\Xi_c(2815)^{+,0}D_s^-$           &$6.80\pm0.40\ (7.30\pm0.40)$    &$-0.90\pm0.07$
&$4.34^{+7.54}_{-3.25}\ (4.65^{+8.08}_{-3.48})$\\
$\Xi_b^{0,-}\to\Xi_c(2815)^{+,0}D_s^{\ast-}$     &$3.27\pm0.015\ (3.47\pm0.017)$    &$-0.41\pm0.04$
&$3.51^{+4.30}_{-2.18}\ (3.74^{+4.58}_{-2.32})$\\
\bottomrule[0.5pt]
\bottomrule[1pt]
\end{tabular*}
\end{table*}

By substituting our numerical results of the form factors and the decay parameters into Eq.~\eqref{eq:physicsP} and Eq.~\eqref{eq:physicsV}, the branching ratios and asymmetry parameters can be further obtained, which are collected in Tables~\ref{tab:nonleptonLambdab}-\ref{tab:nonleptonXib} for the $\Lambda_b^{0}\to\Lambda_c(2625)^{+}M^{-}$ and $\Xi_b^{0,-}\to\Xi_c^{+,0}(2815)M^{-}$ decays, respectively with emitting a pseudoscalar meson ($\pi^{-}$, $K^{-}$, $D^{-}$, and $D_s^{-}$) or a vector meson ($\rho^{-}$, $K^{*-}$, $D^{*-}$, and $D_s^{*-}$). Our results show the process emitted a $\pi^-$, $\rho^-$, or $D_{s}^{(*)-}$ meson has considerable branching ratio, which is possible to be explored in the future experiment, like LHCb. As for other processes, the branching ratios are suppressed by an order of magnitude due to the smaller values of CKM matrix element.

In experiment, the LHCb Collaboration measured~\cite{ParticleDataGroup:2022pth,LHCb:2011poy}
\begin{equation*}
\begin{split}
\mathcal{B}&(\Lambda_b\to\Lambda_c(2625)\pi^-,\Lambda_c(2625)\to\Lambda_c\pi^+\pi^-)\\
&=(3.3\pm1.3)\times10^{-4}.
\end{split}
\end{equation*}
Based on the narrow-width approximation and $\mathcal{B}(\Lambda_c(2625)\to\Lambda\pi^+\pi^-)\approx 67\%$~\cite{ParticleDataGroup:2022pth}, we have $\mathcal{B}(\Lambda_b\to\Lambda_c(2625)\pi^-)=(4.9\pm1.9)\times10^{-4}$, which is apparently smaller than our result. It should be clarified by more precise measurement in future.

In addition, we also compare our results with that in Ref.~\cite{Chua:2019yqh} as shown in the fourth column of Table~\ref{tab:nonleptonLambdab} and Table~\ref{tab:nonleptonXib} for the $\mathcal{B}(\Lambda_b^{0}\to\Lambda_c^{+}(2625)M^-)$ and $\mathcal{B}(\Xi_b^{0,-}\to\Xi_c^{+,0}(2815)M^-)$ decays, respectively. Our results are consistent with the results in  Ref.~\cite{Chua:2019yqh}, but have smaller uncertainties. It benefits from our improved treatment of the baryon wave function. By hypothesizing the $\Lambda_c(2625)$ and $\Xi_c(2815)$ as the dynamically generated resonances from the vector meson-baryon interactions, the authors of Refs.~\cite{Liang:2016ydj,Pavao:2017cpt} calculated the $\Lambda_b\to\Lambda_c(2625)D_s^{-}$ and $\Xi_b\to\Xi_c(2815)\pi^-$ channels. Their results show apparently smaller widths compared with the results from the present work and Ref.~\cite{Chua:2019yqh} based on the $udc$-scheme of the $\Lambda_c(2625)$ and $\Xi_c(2815)$ states. So we also expect the LHCb Collaboration to measure the corresponding $\pi^-$ and $D_s^{-}$ channels, which not only is useful to reveal the inner structures of the $\Lambda_c(2625)$ and $\Xi_c(2815)$, but also can enrich the observed modes of  $b$-decay.

\section{Discussion and conclusion}
\label{sec5}

With the update of High Luminosity Large Hadron Collider and the accumulation of experimental data, the exploration of the bottom baryon decays into the $P$-wave excited charmed baryon becomes highlight. In this work, we study the form factors of the $\Lambda_b\to\Lambda_c(2625)$ and $\Xi_b\to\Xi_c(2815)$ transitions, and further discuss the corresponding semileptonic decays and color-allowed two-body nonleptonic decays.

As the first step, the weak transition form factors are obtained via three-body LFQM, where the important inputs, the spatial wave functions of these concerned baryons, are extracted by solving the Schr\"{o}dinger equation with the support of GEM~\cite{Hiyama:2003cu,Hiyama:2012sma,Yoshida:2015tia} and by adopting a semirelativistic three-body potential model~\cite{Capstick:1985xss,Li:2021qod,Li:2021kfb,Li:2022nim}. By fitting the mass spectrum of the single bottom and single charmed baryons, the parameters in semirelativistic potential model can be fixed. This treatment is different from taking a simple harmonic oscillator wave function with a phenomenal parameter $\beta$. Thus, we can avoid the $\beta$ dependence of the result, where the present work is supported by the baryon spectroscopy.
Additionally, these calculated form factors in this work are comparable with the result from LQCD and consistent with the expectation from HQET.

With the obtained form factors, we further evaluate the weak decays. For the semileptonic processes, our result of $\mathcal{B}(\Lambda_b^0\to\Lambda_c^+(2625)\mu^-\nu_{\mu})=(1.641\pm0.113)\%$ is consistent with current experimental data, and the branching ratios of electron and muon channels can reach up to the magnitude of $1\%$, which are accessible  at the LHCb experiment in future. Besides, other important physical observables, including the leptonic forward-backward asymmetry ($A_{FB}$), the final hadron polarization ($P_{B}$), and the lepton polarization ($P_{\ell}$) are also investigated. As for the nonleptonic processes, the $\pi^-$, $\rho^-$, and $D_s^{(*)-}$-emitted channels have considerable widths, and they are worthy to be focused by LHCb.

{In this work, our study shows the $\Lambda_b\to \Lambda_c(2625)$ and $\Xi_b\to \Xi_c(2815)$ weak transitions have sizable branching ratios, which can be accessible at experiment. Especially, we notice that different theoretical groups gave different results of these discussed transitions by different theoretical frameworks and different structure assignments to the $\Lambda_c(2625)$ and $\Xi_c(2815)$ \cite{Pervin:2005ve,Liang:2016ydj,Pavao:2017cpt,Gutsche:2018nks,Chua:2019yqh,Nieves:2019kdh}, which can be tested by future experimental measurement. At present, only the $\Lambda_b^0\to \Lambda_c^+(2625)\mu^-\nu_\mu$ was measured \cite{ParticleDataGroup:2022pth}. Considering the high-luminosity upgrade to LHC, the LHCb experiment will have enough interest and potential to carry our the measurement to these discussed weak transitions in this work. Taking this opportunity, we suggest LHCb to measure these discussed channels, where these measurements no doubt can be useful to enrich the $b$-decay modes and can of course be applied to distinguish different structure assignments to the $\Lambda_c(2625)$ and $\Xi_c(2815)$ states.}

\section*{ACKNOWLEDGMENTS}

This work is supported by the China National Funds for Distinguished Young Scientists under Grant No. 11825503, National Key Research and Development Program of China under Contract No. 2020YFA0406400, the 111 Project under Grant No. B20063, the National Natural Science Foundation of China under Grant No. 12247101, the project for top-notch innovative talents of Gansu province, and by the Fundamental Research Funds for the Central Universities under Grant No. lzujbky-2022-it17.

\end{document}